\documentclass[twocolumn,amssymb,amsmath,superscriptaddress,aps,pre]{revtex4-1}
\usepackage{epsfig}
\usepackage{times}
\usepackage{bm}

\begin{document}

\title{Activation process in excitable systems with multiple noise sources: One and two interacting units}

\author{Igor Franovi\'c}
\email{franovic@ipb.ac.rs}
\affiliation{Scientific Computing Laboratory, Institute of Physics, University of Belgrade, P.O. Box 68,
11080 Beograd-Zemun, Serbia}

\author{Kristina Todorovi\'c}
\affiliation{Department of Physics and Mathematics, Faculty of Pharmacy, University of Belgrade,
Vojvode Stepe 450, Belgrade, Serbia}

\author{Matja\v{z} Perc}
\email{matjaz.perc@uni-mb.si}
\affiliation{Faculty of Natural Sciences and Mathematics, University of Maribor,\\
Koro\v{s}ka Cesta 160, SI-2000 Maribor, Slovenia}
\affiliation{Department of Physics, Faculty of Sciences, King Abdulaziz University, Jeddah, Saudi Arabia}

\author{Neboj\v{s}a Vasovi\'c}
\affiliation{Department of Applied Mathematics, Faculty of Mining and Geology, University of Belgrade, P.O. Box 162, Belgrade, Serbia}

\author{Nikola Buri\'c}
\email{buric@ipb.ac.rs}
\thanks{corresponding author}
\affiliation{Scientific Computing Laboratory, Institute of Physics, University of Beograd, P.O. Box 68, 11080 Beograd-Zemun, Serbia}

\begin{abstract}
We consider the coaction of two distinct noise sources on the activation process of a single and two interacting excitable units, which are mathematically described by the Fitzhugh-Nagumo equations. We determine the most probable activation paths around which the corresponding stochastic trajectories are clustered. The key point lies in introducing appropriate boundary conditions that are relevant for a class II excitable unit, which can be immediately generalized also to scenarios involving two coupled units. We analyze the effects of the two noise sources on the statistical features of the activation process, in particular demonstrating how these are modified due to the linear/nonlinear form of interactions. Universal properties of the activation process are qualitatively discussed in the light of a stochastic bifurcation that underlies the transition from a stochastically stable fixed point to continuous oscillations.
\end{abstract}


\maketitle

\section{Introduction}

Excitability is a dynamical feature shared by nonlinear systems that lie in vicinity of bifurcation underlying transition from stationary state toward the sustained periodic activity \cite{I07}. Study of models where stochastic excitable dynamics is crucial for shaping local or collective behavior is a rapidly developing field, gaining relevance in terms of theory as well as for offering qualitative insights into a variety of biological \cite{I07,WRK00,CLC11} and inorganic systems \cite{YMRBSRL06,LHMY02,BYSRBL12}.

Excitability is manifested in the existence of a narrow range of stimuli magnitudes where marginally different perturbations may cause the system to generate two qualitatively distinct types of responses. While smaller perturbations give rise to small-amplitude (linear) responses, the slightly larger perturbations may elicit pulse-like, large-amplitude excitation loops. The latter are comprised of activation and relaxation stage, whereby the spike profile remains independent on the form of applied perturbation. If one interprets perturbation
in terms of setting the particular initial conditions, then the excitability feature is reflected in the point that the system shows strong sensitivity to initial conditions within a small domain of relevant values. Consequently, behavior of excitable systems is heavily susceptible to noise \cite{LGNS04}, whose influence may at least in part be understood as excitability amplification.

Given the strong likelihood of encountering systems which are influenced by combined action of variations in the environment and the fluctuations of the internal parameters, it is justified to analyze models where different sources of noise affect the dynamics of multiple variables. This point, together with the fact that the existence of pulse-like excitations requires a flow with dimension no less than two \cite{W00}, suggests that the system with two variables, each subjected to stochastic perturbation, may be considered a paradigmatic setup. For example, two sharply separated characteristic time scales and two sources of noise acting independently on the fast and the slow variable are relevant ingredients for the description of a typical neuron \cite{DRL12,SJ02} or laser dynamics \cite{YMRBSRL06}. For neurons, stochastic term affecting the fast variable may account for synaptic noise due to random arrival of spikes from a large number of afferents, whereas the stochastic component in the slow variable dynamics may be seen as internal noise, associated to thermal fluctuations in the opening of the ion-gating channels. Consistent with this, we refer to stochastic term added to the dynamics of the fast (slow) variable as external (internal) noise.

From the theoretical viewpoint, a comprehensive insight into the noise-driven activation process is fundamental for understanding excitability. The problem we focus on concerns the activation process in a single or two coupled excitable Fitzhugh-Nagumo ($FHN$) elements driven by two independent sources of noise. The main reason for considering the $FHN$ model lies in its generic character, viz. the fact that it is canonical for class II excitable systems, which involve an almost continuous transition between the small- and the large-amplitude excitations. For conceptual reasons and in view of potential applications, only the spiking responses are relevant and of interest to us. It follows that the formulation of activation event has to be adapted to class II excitable systems, such that it satisfies two criteria: $(i)$ it should warrant a clear-cut distinction of the spiking response from the small-amplitude excitation and $(ii)$ it should allow an immediate generalization from the case of a single unit to that of two interacting units. We stress that in order to meet both criteria, it is crucial to introduce an appropriate terminating boundary set for the activation problem, as demonstrated in the paper.

Note that the term activation is inherited from, but it is not used in the same sense as in a typical escape problem \cite{LGNS04,HE05,MS92,MS96,DLMS96}. To explain the differences, one first invokes a general remark that for excitable systems, two possible types of response derive from the proximity to bifurcation rather than the bistable dynamics, so that the threshold-like behavior is not associated to a genuine separatrix (saddle structure), but
rests on the coaction of nonlinearity and the sharp separation between the system's characteristic time scales.
Thus, resolving the origin and character of threshold-like behavior for excitable elements may be linked to an unstable fixed point or, as in case of a $FHN$ unit, to a quasi-separatrix ("ghost separatrix") \cite{KPLM13}. This is apparently different from the typical escape problem scenario.

Another important point is that we tie the activation exclusively to spiking response, so that the event where the phase point reaches the quasi-separatrix \emph{per se} is insufficient to count as activation. In other words, crossing the quasi-separatrix does not present a discriminative condition between the small- and the large-amplitude excitations. This is why we introduce novel terminating boundary conditions, rather than just consider an extension of the escape problem to excitable systems \cite{KPLM13}. One should also point out that our approach is conceptually different from the earlier work concerning noisy excitable neurons \cite{PPS05,PPM05}, where the terminating boundary does not constitute a set, but rather a unique boundary point, introduced as an arbitrary threshold independent on the structure of phase space.

We analyze the activation problem from two angles, one by determining the most probable activation paths ($MPAP$s), and the other by examining the statistical features of the activation process. An important result consists in determining the $MPAP$s for a single and two coupled units under different ratios of external vs internal noise $D_1/D_2$. In case of two units, the topology of the obtained trajectories is shown to qualitatively depend on the form of coupling. 
The statistics of a single- and two-unit activation process is characterized by examining how the
time-to-first-pulse ($TFP$) $\tau$ averaged over different stochastic realizations and the associated  coefficient of variation $R$ depend on noise. Both $\tau(D_1, D_2)$ and $R(D_1, D_2)$ are found to display universal behavior for all considered setups, whereby the transition between two of the observed $\tau$ regimes is associated to the fact that a unit undergoes stochastic bifurcation induced by $D_1$ and $D_2$. For two units, the form of coupling is shown to have a nontrivial effect on correlation of the individual mean $TFP$s, which we relate to synchronization properties of the time series for the given parameter set. 

The paper is organized as follows. In Sec. \ref{Model}, we present the details of the model, focusing on the threshold-like behavior of a single unit and the results of bifurcation analysis for the coupled units. Section \ref{MPAP} concerns the case of a single unit subjected to external and internal noise. Having laid out the
details of the method used to determine the $MPAP$s, cf. subsection \ref{Method}, we examine how the topological features of the $MPAP$s depend on the pertaining noise intensities. Apart from relating the properties of $\tau(D_1, D_2)$ dependence to the onset of stochastic bifurcation from the stochastically stable fixed point to the stochastically stable limit cycle, we also introduce an approximation to explicitly demonstrate that $D_1$ and $D_2$ make substantially different impact on the mean $TFP$s. Section \ref{Twounits} contains the results for two units interacting via the linear or nonlinear couplings. It is analyzed how the different form of coupling affects the profile of the respective $MPAP$s, the stochastic bifurcation as well as the correlation of single unit mean $TFP$s. Section \ref{Conc} provides a summary of the main results.

\section{Background on the applied model}
\label{Model}

\subsection{Dynamics of a single excitable unit}
\label{Single}

As a paradigm for excitable systems, we consider the $FHN$ model, so that the dynamics of a single unit is given by
\begin{align}
dx&=f_x(x,y)=[x-x^3/3-y]dt+\sqrt{2D_1}dW_1\nonumber\\
dy&=f_y(x,y)=\epsilon (x+b)dt+\sqrt{2D_2}dW_2. \label{eq1}
\end{align}
The model is canonical for type II excitability class, meaning that the equilibrium lies in vicinity of the direct supercritical Hopf bifurcation. The latter is controlled by the excitability parameter $b$, whereby the critical value is $|b|=1$. For $|b|>1$, the system possesses a unique stable equilibrium $(x_{eq},y_{eq})=(-b,-b+b^3/3)$, whereas for $|b|<1$ the oscillatory state sets in. Given the symmetry of the system \eqref{eq1}, the analysis may be confined to case $b>0$ without loss of generality. The unit in subcritical state displays excitable behavior if $b$ is kept close to bifurcation threshold. In this paper, we fix $b=1.05$.

The other important ingredient of the model is the sharp separation between the characteristic time scales of the activator and the recovery variable. The fast-slow dynamics is facilitated by setting $\epsilon$ to a small value ($\epsilon=0.05$). So far, $FHN$ model has been applied in describing the dynamics of electrochemical reactions \cite{KWS98} and cardiac cells \cite{CLC11}, but is best known for its role in the field of neuroscience \cite{I07,LGNS04}. In the latter context, the fast variable may be viewed as analogous to the neuron membrane potential, whereas the action of the slow variable may qualitatively be compared to that of $K^+$ ion-gating channels \cite{I07}. Regarding the impact of random perturbations, we are interested into how the activation of units is shaped by two independent sources of noise. In \eqref{eq1}, the stochastic effects are represented by the Wiener processes whose increments satisfy $\langle dW_i \rangle=0$ and $\langle dW_idW_j\rangle=dt\delta_{ij}$ for $i,j\in{1,2}$. The dynamics of an excitable unit may be summarized as follows. In absence of perturbation, the selected parameter values are such that the system lies at equilibrium. Kicked by the perturbation, the unit may either display a small amplitude response, whereby the phase point rapidly decays back to equilibrium, or may exhibit a large excursion, settling to equilibrium only after the phase point has traversed the orbit corresponding to a complete oscillation cycle.

\begin{figure}
\centerline{\epsfig{file=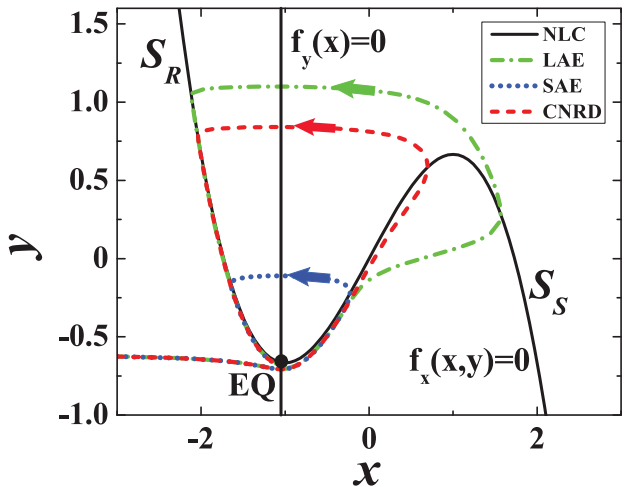,width=8.5cm}}
\caption{(Color online) Phase plane analysis for a $FHN$ excitable unit. Equilibrium ($EQ$) lies at the intersection of the nullclines $f_x(x,y)=0$ and $f_y(x)=0$. The $x$-nullcline comprises three branches. In the singular limit $\epsilon\rightarrow0$, the spiking branch $S_S$ and the refractory branch $S_R$ are attractive, whereas the orbit separating the initial conditions that lead to small- or large-amplitude excitations, $SAE$ and $LAE$ respectively, contains the middle unstable branch. For finite $\epsilon$, the boundary between the two sets of initial conditions foliates into a thin layer of canard-like trajectories ($CNRD$) referred to as "ghost separatrix". Illustrated trajectories are obtained by fixing the $x$ initial condition to $x_0=-3$, whereas $y_0$ is varied. Extreme sensitivity to initial conditions in vicinity of ghost separatrix is corroborated by the fact that a difference in $y_0$ of the order of $10^{-14}$ gives rise to a different form of excitation. The parameters of $FHN$ model are fixed to $\epsilon=0.05,b=1.05$.}
\label{Fig1}
\end{figure}

For the proper statement of our problem, it is important to first consider the geometric interpretation of the unit's dynamics. Two particular issues are addressed, one related to the existence and the structure of the boundary between the initial conditions resulting in small- or large-amplitude responses, whereas the other concerns the distinction between the roles of $D_1$ and $D_2$ in triggering the pulse emission. Regarding the first point, we summarize the results on the deterministic (noiseless) version of the system \eqref{eq1} obtained by the method of phase plane analysis, which rests on the singular perturbation theory. In the limit $\epsilon\rightarrow0$, \eqref{eq1} turns into a one-dimensional system $\dot{x}=f_x(x,y)$ under the constraint $\dot{y}=0$, meaning that $y$ may be viewed as a fixed parameter. For small but finite $\epsilon$, $x$ quickly relaxes to the value given by $f_x(x,y)=0$, the condition outlining the curve referred to as the $x$ (cubic) nullcline or the slow manifold, see Fig. \ref{Fig1}. The $x$ nullcline is comprised of three branches, whose stability features derive from the singular limit $\epsilon\rightarrow0$.  While the refractory $S_R$ and the spiking branch $S_S$ may then be regarded as attractors, the middle branch is unstable and is a part of separatrix between the attractive branches. The key point is that for small but finite $\epsilon$, the structure of the boundary and the related threshold behavior is mostly inherited from the singular limit. The distinction is that finite $\epsilon$ induces foliation of the boundary around the maximum of the $x$-nullcline. What has explicitly been shown by the so-called "blow-up method" \cite{KS01}, is that the boundary between the initial conditions leading to $S_S$ or $S_R$ is not given by a single line, but rather by a thin layer made up of an infinite family of canard-like trajectories of system \eqref{eq1}. The latter further implies that the boundary constitutes an invariant set. Boundary layer may still be perceived as a single line, a kind of "ghost separatrix" \cite{KPLM13}, because at distances $d>>\epsilon$ from the fold point $(1,2/3)$, all the constituent trajectories become virtually indistinguishable.

As for the qualitative interpretation of the effects of noise, it is evident that the noise term added to the slow-variable dynamics may shift the position of the $y$-nullcline. In other words, internal noise is capable of translating the fixed point from the stable refractory to the unstable (middle) branch of the $x-$ nullcline, which temporarily switches the system from excitable to oscillatory state. While the impact of $D_2$ can be understood by geometric analysis, there is no analogous interpretation in case of $D_1$.

\subsection{Dynamics of a couple of excitable units}
\label{Couple}

In case of a pair of coupled units, we consider two distinct setups, one where the units interact via the symmetrical linear couplings, and the other involving couplings given by the nonlinear threshold-like function.

The dynamics of a couple of $FHN$ units interacting via linear couplings is given by
\begin{align}
dx_i&=[x_i-x_i^3/3-y_i]dt+\sqrt{2D_1}dW_1^i+c[x_i-x_j]dt \nonumber\\
dy_i&=\epsilon (x_i+b)dt+\sqrt{2D_2}dW_2^i,  \label{eq2}
\end{align}
where $i,j\in\{1,2\},i \neq j$ specify the individual units. The random perturbations are such that the terms acting on different elements are supposed to be uncorrelated $\langle dW_k^idW_l^j\rangle=0, k,l\in{1,2}$.
Regarding the system parameters, note that the units are assumed to be identical (same $b$ and $\epsilon$), while the symmetrical couplings are characterized by the coupling strength $c$. To see how the unit's excitability feature is modified in presence of interaction, we make a brief summary of the results of the bifurcation analysis carried out on the noise-free version of the system \eqref{eq2}. The first remark is that the equilibrium is located at $(x_1,y_1,x_2,y_2)=(-b,-b+b^3/3,-b,-b+b^3/3)$, whereby its stability is determined by the two pairs of complex conjugate characteristic exponents which satisfy $\mu_{1,2}=[1-b^2+2c\pm\sqrt{(1-b^2+2c)^2-4\epsilon}]/2$ and
$\mu_{3,4}=[1-b^2\pm\sqrt{(1-b^2)^2-4\epsilon}]/2$. Given that the single unit parameters are kept fixed, the local stability of equilibrium is changed under variation of the coupling strength. In particular, it may be demonstrated that the system undergoes a direct supercritical Hopf bifurcation at the critical value $c_{H,L}=(b^2-1)/2$. At this point, the stability of equilibrium is lost and the units are no longer in the excitable regime. However, we stress that the complete picture on the system dynamics cannot be gained from the local bifurcation analysis alone, because even before the excitability feature is lost (for $c=c_{H,L}$), it is modified due to a global fold-cycle bifurcation which occurs at $c_{FC}<c_{H,L}$. Thus, under increasing $c$ the system exhibits three types of characteristic behavior: $(i)$ the "proper" excitable regime for $c<c_{FC}$, $(ii)$ the regime of "generalized excitability" for $c\in(c_{FC},c_{H,L})$ and $(iii)$ the oscillatory state at $c>c_{H,L}$. The case $(ii)$ involves coexistence between the fixed point and the large limit cycle, whose basins of attraction are separated by the stable manifold of the saddle cycle. In strict terms, such a scenario does not conform to Izhikevich's definition of excitability, because instead of relaxing to equilibrium after the large excursion, the system displays continuous oscillations. Nevertheless, some authors consider such a behavior as excitable or excitable-like, and one should also note that the oscillation orbit may still pass quite close to equilibrium, depending on the position, size and the manifold structure of the saddle cycle. Note that above $c_{H,L}$, the large limit cycle created in the global event survives as the only attractor, because the incipient cycle born via Hopf bifurcation grows only until colliding with the preexisting saddle-cycle, such that the two get annihilated in the inverse fold-cycle bifurcation. In other words, the properties of the limit cycle attractor both below and above $c_{H,L}$ are determined by the global bifurcation. 

Having summarized the results of the bifurcation analysis for the setup involving two units coupled via linear function, we turn to the scenario where the units subjected to external and internal noise interact in a nonlinear fashion. The dynamics of the units then obeys
\begin{align}
dx_i&=[x_i-x_i^3/3-y_i]dt+\sqrt{2D_1}dW_1^i+c\arctan(x_j+b)dt \nonumber\\
dy_i&=\epsilon (x_i+b)dt+\sqrt{2D_2}dW_2^i,  \label{eq3}
\end{align}
The interaction terms have a form of a threshold-like function, whose argument is defined relative to the corresponding unit's equilibrium $x_j-x_{j,EQ}=x_j+b$. This way, the impact of the state $x_j$ is felt stronger if it lies further away from the equilibrium. The stability of equilibrium is determined by the four roots of the characteristic equation, which appear as two pairs of complex conjugates $\mu_{1,2}=[1-b^2+c\pm\sqrt{(1-b^2+c)^2-4\epsilon}]/2$, $\mu_{3,4}=[1-b^2-c\pm\sqrt{(1-b^2-c)^2-4\epsilon}]/2$. Again, it may be shown that the system \eqref{eq3} undergoes a direct supercritical Hopf bifurcation at $c_{H,NL}=b^2-1$. Note that we confine the analysis to the case $c>0$. Unlike the setup based on the linear coupling, here one does not encounter the global bifurcation controlled by $c$. In other words, the equilibrium is stable and the units are in excitable regime for $c<c_{H,NL}$, whereas the system is in oscillatory state for $c>c_{H,NL}$. 

\section{Activation process in an excitable unit driven by external and internal noise} \label{MPAP}

In the following subsection we introduce the numerical method applied to determine the $MPAP$s. The method is illustrated in case of a single $FHN$ unit, but it can readily carry over to the process of first pulse emission for two interacting units. In subsections \ref{Examples} and \ref{Twopaths}, the topological features of the pertaining trajectories will be analyzed in reference to stochastic bifurcation, viz. for noise intensities substantially below, near or above the critical domain of $(D_1,D_2)$ values. While the main focus lies with the $MPAP$s explicitly obtained from stochastic simulations, the extended discussion will also concern the trajectories generated as solutions of the effective Hamiltonian equations under boundary conditions relevant for the process of first pulse emission. We stress that the use of such equations is distinct from the standard Hamiltonian approach which yields optimal trajectories in a typical escape problem, or the generalization of an escape problem to an excitable $FHN$ unit. In order to set up the discussion carried out in subsections \ref{Examples} and \ref{Twopaths}, the following subsection addresses the precise role of these effective equations and the ensuing trajectories.

\subsection{Method applied to determine the $MPAP$s} \label{Method}

The problem of obtaining the $MPAP$s for excitable units in general comprises two issues. One issue concerns how the terminating boundary conditions are specified, whereas the other relates to the particular details of the method.  A common ingredient in previous approaches to analysis of activation process in excitable units has been to draw an analogy to motion of a particle in a one-dimensional potential perturbed by noise. Reduction to a one-dimensional problem naturally utilizes the decomposition of the system dynamics to fast and slow motions. Though such approximate methods are not intended to trace the unit's most likely activation paths, the use of Fokker-Planck formalism still allows one to gain insight into certain statistical features of the activation process, including the mean activation time and its variance \cite{HE05,BR11}. Nevertheless, an inherent drawback consists in the lack of ability to account for the simultaneous influence of perturbations added to both the slow and the fast component. In conceptual terms, the key problem lies in the fashion by which the escape from the stationary state is precisely defined. In particular, instead of associating the terminating boundary to the structure of the system's phase space, the activation event is considered as a crossing of a predefined threshold, typically coinciding with the fold point at the minimum of the slow manifold. Compared to such methods, the approach we apply is preferred because $(i)$ it introduces a unique definition of the activation path consistent with the structure of the phase space, $(ii)$ one may consider the coaction of random perturbations added to both the fast and the slow subsystem, and $(iii)$, one is able to explicitly determine the $MPAP$s around which the different stochastic realizations are clustered.

Before proceeding to the details of the numerical method implemented for calculation of the $MPAP$s, let us specify the boundary conditions relevant for the problem of first pulse emission in case of an excitable $FHN$ unit. In particular, for the noise-driven system \eqref{eq1}, we consider the stochastic fluctuation paths in configuration space $(x,y)$ that emanate from the deterministic fixed point $(x_{eq},y_{eq})=(-b,-b+b^3/3)$ and terminate at the spiking branch of the cubic ($(x)$) nullcline. Given selection of the terminating boundary set derives from the fact that the spiking branch of the cubic nullcline defines the spike profile for the deterministic limit cycle in the superthreshold regime $b\lesssim1$, while the analogous point holds for the noise-induced oscillations in the excitable regime $b\gtrsim1$. In these terms, it has been verified that the profiles of spikes for an arbitrary combination of relevant $(D_1,D_2)$ values can be approximated with sufficient accuracy by the orbit corresponding to the deterministic limit cycle characterizing the supercritical state. Note that the terminating boundary set we introduce is distinct from the one in \cite{KPLM13}, where a conceptually distinct problem has been considered. In particular, the aim in that paper has been to extend the standard escape problem to the case of an excitable $FHN$ unit, which has been achieved by fixing the "ghost separatrix" as the terminating boundary set. The advantage of the boundary conditions we have adopted is that they can readily be generalized to the case of two interacting excitable units. In another paper, we shall further demonstrate that the relevance of a proper problem formulation which warrants that the small-amplitude response and the spiking response are clearly distinguished, as well as the related selection of the appropriate terminating boundary conditions, is even more pronounced when analyzing the noise-driven first pulse emission process for an assembly of excitable units.

The details of the numerical method applied to determine the $MPAP$s are as follows. For the given $(D_1,D_2)$, we consider an ensemble of fluctuation paths $(x(t),y(t))$ which start from the deterministic fixed point $(x_{eq},y_{eq})$ at moment $t_i$ and satisfy the above stated terminating boundary conditions. The terminating time $t_f$, as well as the associated coordinates $(x(t_f),y(t_f))$ are left unspecified. For the described ensemble, we study the statistics of the $(x(t),y(t))$ position of trajectories as a function of time $t_i<t<t_f$ preceding the arrival to the terminating boundary set. In general, the idea is to sample the different stochastic realizations of activation trajectories in order to determine histograms of the path history reaching the terminating boundary set. Naturally, the recorded trajectories are characterized by the different $t_f$ times. Thus, the proper approach to characterize the statistics of the paths in configuration space is to consider the prehistory probability density \cite{DMSSS92}, which concerns the distribution of paths ending at the specified boundary set. To obtain the former, one effectively sets the time when each stochastic realization terminates to $t=0$, such that the behavior of the process during the initiation of the pulse is observed by looking backward in time. This approach has been introduced in \cite{DMSSS92}, and has been applied a number of times since \cite{NBK13}. The prehistory probability distribution is defined as
\begin{align}
H(x,y,t)dxdy&=Pr[x(t)\in(x,x+dx),y(t)\in(y,y+dy)| \nonumber \\
&x_b(t_f),y_b(t_f),x(t_i)=x_{eq},y(t_i)=y_{eq}], \nonumber \\
& t_i<t<t_f, x<x_b(t_f),y<y_b(t_f). \label{eq4}
\end{align}
The most probable path for the first pulse emission process is determined by collecting the points $(x_m(t),y_m(t))$ which correspond to the maximum of $H(x,y,t)$ at any given moment $t$. In other words, the $MPAP$ for the given $(D_1,D_2)$ by definition coincides with the peak of $H$ as a function of time. In practice, the numerical method used to determine the prehistory probability density has involved dividing the $(x,y)$ phase space into a grid of $70$ by $70$ cells of length $\Delta x=0.048$ and width $\Delta y=0.09$. Throughout the paper, the numerical integration of the appropriate set of stochastic equations is carried out by the Heun algorithm with the time step $\delta t=0.002$, whereas the averaging is performed over an ensemble of $5000$ different stochastic realizations of the first pulse emission process.

In order to provide qualitative guidelines for an extended discussion, the $MPAP$s determined via the above described method for different setups and the different domains of noise intensities will be compared to trajectories obtained by integrating the set of effective Hamiltonian equations under boundary conditions relevant for the problem of first pulse emission. One recalls that in case of a typical escape problem, the Hamiltonian theory formulated in the extended variable-momentum state space may be used to explicitly obtain the optimal trajectories which coincide with the $MPAP$, whereby such trajectories are determined by the minimum of action, introduced as an effective cost function. In \cite{KPLM13}, such an approach has been implemented for an extension of the escape problem to an excitable $FHN$ unit, having considered the ghost separatrix as the terminating boundary. Due to fundamentally different terminating boundary conditions, one cannot apply the Hamiltonian theory \emph{per se} for the problem of first pulse emission, as we explain in more detail further below. Therefore, our approach cannot be interpreted in the genuine context of or be referred to as an extension of the Hamiltonian theory. In fact, the Hamiltonian system we consider should be interpreted as a set of effective equations integrated for relevant boundary conditions, whereby a particular trajectory is singled out according to a certain predefined recipe. Our current goal is not to provide a systematic theory, but only to point out to striking similarity between the numerically obtained $MPAP$s for the process of first pulse emission and the trajectories generated by the set of effective Hamiltonian equations. This can be considered a preliminary stage of a study which may ultimately lead to derivation of an appropriate theory for the process of first pulse emission, whose basis would take into account certain aspects of the standard Hamiltonian formalism.

In the remainder of this subsection, we briefly consider the main elements of the Hamiltonian approach, whose detailed description may be found in \cite{BMLSM05}, and clarify the differences emerging in case of the first pulse emission process. Within the Hamiltonian formalism, the optimal trajectories for a noise-driven escape process are obtained by variation-like approach which minimizes the appropriately defined "cost functional" $\bar{S}(\mathcal{J},t)$ along the set of possible activation trajectories $\mathcal{J}$ between the two fixed boundaries. In particular, the probability for noise to induce the $x_i\mapsto x_f$ transition is given by
$p(x_f|x_i)=\int_\mathcal{J}P[\mathcal{J}]d\{\mathcal{J}\}$, such that $d\{\mathcal{J}\}$ denotes  integration along all the paths $\{\mathcal{J}=\{x_1,\dots,x_N\}\}$ connecting $x_i$ and $x_f$. Each path is weighted by the probability $P[\mathcal{J}]\propto\exp[\frac{-S(x_1,\dots,x_N)}{D}]$, where $S$ is the cost function for the particular path. When $D$ is small, the largest contribution to $p(x_f|x_i)$ comes from the path with the minimal cost function
$S_{min}=\text{min}\{S_{\mathcal{J}}|\mathcal{J}=\{x_1,\dots,x_N\}\}$. In other words, small noise rarely gives rise to transition events, but once they occur, all the other orbits are suppressed in favor of the one corresponding to the minimal cost function. The transition probability then takes the asymptotic form
\begin{align}
p(x_f|x_i)= z\exp[\frac{-S_{min}}{D}],x_1\equiv x_i,x_N\equiv x_f, \label{eq5}
\end{align}
whereby the prefactor $z$ is associated to "degeneracy" of the minimum, or rather the number of trajectories close to the one given by $S_{min}$ \cite{BMLSM05}.

The form of $\bar{S}$ is selected so to reflect on one hand the impact of noise, whereas on the other hand to explicitly incorporate the constraints between the variables and the stochastic terms which derive from the equations of motion. The necessary conditions for the minimum of such a constraint problem can then be treated by the Lagrange multiplier technique. Ultimately, one arrives at a set of equations that may be interpreted as a Hamiltonian system, where the Lagrange multiplier $\lambda$ enact the momenta conjugate to the system variables. In particular, for system \eqref{eq1}, the corresponding Hamiltonian set including the generalized momenta is given by
\begin{align}
\frac{dx}{dt}&=x-x^3/3-y+r_xp_x \nonumber\\
\frac{dy}{dt}&=\epsilon(x+b)+r_yp_y \nonumber\\
\frac{dp_x}{dt}&=-(1-x^2)p_x-\epsilon p_y \nonumber\\
\frac{dp_y}{dt}&=p_x, \label{eq6}
\end{align}
where $p_x$ and $p_y$ are the components of the momentum, while $r_x$ and $r_y$ represent the scaled noise intensities. By the latter notation, $D_1$ and $D_2$ are conveniently expressed in terms of a single
amplitude $D$ ($D_1=r_xD$ and $D_2=r_yD$). System \eqref{eq6} has the Wentzel-Freidlin Hamiltonian $H=p_x(x-x^3/3-y)+p_y\epsilon(x+b)+\frac{r_x}{2}p_x^2+\frac{r_y}{2}p_y^2$, whereas the trajectories connecting
the initial state $i$ and final state $f$ are characterized by the action $S=\int_{t_i}^{t_f}dt\frac{1}{2}(r_xp_x^2+r_yp_y^2)$. At variance with system \eqref{eq1}, the fixed point of the system \eqref{eq6}, $(x_0,y_0,p_{x,0},p_{y,0})=(-b,-b+b^3/3,0,0)$, is unstable for $b>1$, which is corroborated by the positive real part of the characteristic exponents $\mu_{1,2}=\frac{-(1-b^2)\pm\sqrt{(1-b^2)^2-4\epsilon}}{2}$.

Conceptually, the Hamiltonian formalism is typically applied to escape problems, where the stable equilibrium coexists with a certain saddle state, most often the saddle cycle \cite{BMLSM05}. The aim is then to obtain the heteroclinic trajectory in the extended space, which emanates from the unstable manifold of the fixed point at $t\rightarrow-\infty$ and reaches the stable manifold of the saddle cycle asymptotically at $t\rightarrow\infty$, as well as tangentially ($\textbf{p}\rightarrow0$) for $t\rightarrow0$. In principle, depending on $\textbf{p}$, the dynamics in the extended space may also support the trajectories that do not settle at, but run across the cycle, as well as the trajectories that are repelled by the cycle, thus being reflected back to initial state. In \cite{KPLM13}, an extension of the escape problem to a single $FHN$ unit has been considered by utilizing the threshold-like behavior to construct the "ghost" manifold (a separatrix in the asymptotic limit
$\epsilon\rightarrow 0$), which plays the role of a saddle structure. For such a scenario, the Hamiltonian approach yields the optimal trajectory connecting the unstable manifold of the fixed point to the stable manifold of the ghost separatrix.

Compared to this, the problem of first pulse emission is fundamentally different because one should look for trajectories that cross the ghost separatrix, viz. leave the basin of attraction of the fixed point. While the action surface in vicinity of the basin boundary should have one clearly defined global minimum, the physical picture near the spiking branch of the $x$-nullcline typically involves multiple local minima of the action surface.
A deeper theoretical analysis of these issues is beyond the scope of the current paper. The point we make here is as follows. Let us consider the trajectories that correspond to local minima of the action surface and adopt as a rule to select the solution that has the smallest momentum at the terminating boundary. Then, for different system configurations, the comparison of the $MPAP$s numerically obtained from an ensemble of stochastic realizations reveals significant similarity to the respective trajectories generated by the set of effective Hamiltonian  equations.

In terms of numerical treatment, obtaining the relevant trajectories from the effective system \eqref{eq6} comprises a boundary value problem. At the initial moment $t_i$, the coordinates $(x(t_i),y(t_i),p_x(t_i),p_y(t_i))$ lying on the unstable manifold of the fixed point are specified according to the method provided in \cite{BMLSM05}, which connects the coordinates in the configuration space with the components of the momentum. By this method, it follows that the initial coordinates in the extended space can be parameterized via a single angular variable $\phi\in(0,2\pi)$. The different trajectories are then obtained by sampling $1000$ equally spaced initial conditions that cover the entire range of $\phi$ values. The moment of arriving at the second boundary $t_s$, as well as the coordinates $(x(t_s),y(t_s),p_x(t_s),p_y(t_s))$ are not explicitly specified. In effect, we apply a shooting approach that involves a set of different trajectories with particular initial conditions, whereby all the trajectories terminate at the spiking branch of the cubic nullcline. System \eqref{eq6} is integrated by a standard solver implementing the fourth-order Runge-Kutta routine. The cost function $S$ is calculated along each of the trajectories, and we single out the trajectory that corresponds to a local minimum having the smallest value of angular momentum at the terminating boundary (typically of the order of $10^{-4}$ or less). Note that the analogous numerical approach is applied in case of two interacting units. The only difference is that the initial conditions are parameterized in terms of three, instead of a single angular variable.

\subsection{Examples of MPAPs and the method's persistence under increasing noise} \label{Examples}

We systematically examined how the $MPAP$s change under variation of the ratio of external vs internal noise intensity $r_x/r_y$. It has been established that the $MPAP$ profiles are not sensitive to gradual changes over the whole range of $r_x/r_y$ values. In particular, when slowly increasing $r_x/r_y$, the trajectories are found to either exhibit barely visible changes in shape, or to converge to each other once the "ghost separatrix" is crossed. Therefore, in qualitative terms one may single out three characteristic forms of solutions, corresponding to cases $(r_x,r_y)=(1,0), (r_x,r_y)=(0,1)$ and $(r_x,r_y)=(1,1)$, which can be held representative for the problem of first pulse emission. In other words, the topological features of $MPAP$s are primarily sensitive to whether a particular noise source or both sources are present in the system \eqref{eq1}.

The other point one has to consider is the impact of the noise intensity $D$. Naturally, the physical picture described above is maintained for sufficiently small $D$. The analysis provided in subsection \ref{Numerics} will show that the term "sufficiently small" $D$ effectively implies that the system lies sufficiently away from the stochastic bifurcation underlying transition from stochastically stable fixed point to the noise-induced oscillations. Above the stochastic bifurcation, the attractive power of the fixed point is no longer felt, such that the noise can move the phase point away from equilibrium without an opposing force. Thus, if one focuses on $MPAP$s for any of the three characteristic $r_x/r_y$ ratios, the impact of noise will become substantial as one approaches the stochastic bifurcation, and will become overwhelming above the stochastic bifurcation. Note that for large $D$, the $MPAP$s determined via the method introduced in subsection \ref{Method} lose physical meaning, because the fluctuations over an ensemble of stochastic realizations grow too large to be accurately described by the maximums of the prehistory probability density $H(x,y,t)$. The influence of noise is reflected in the increase of the skewness and the kurtosis of the distribution of system's variables for different stochastic realizations at arbitrary $t$.

\begin{figure}
\centerline{\epsfig{file=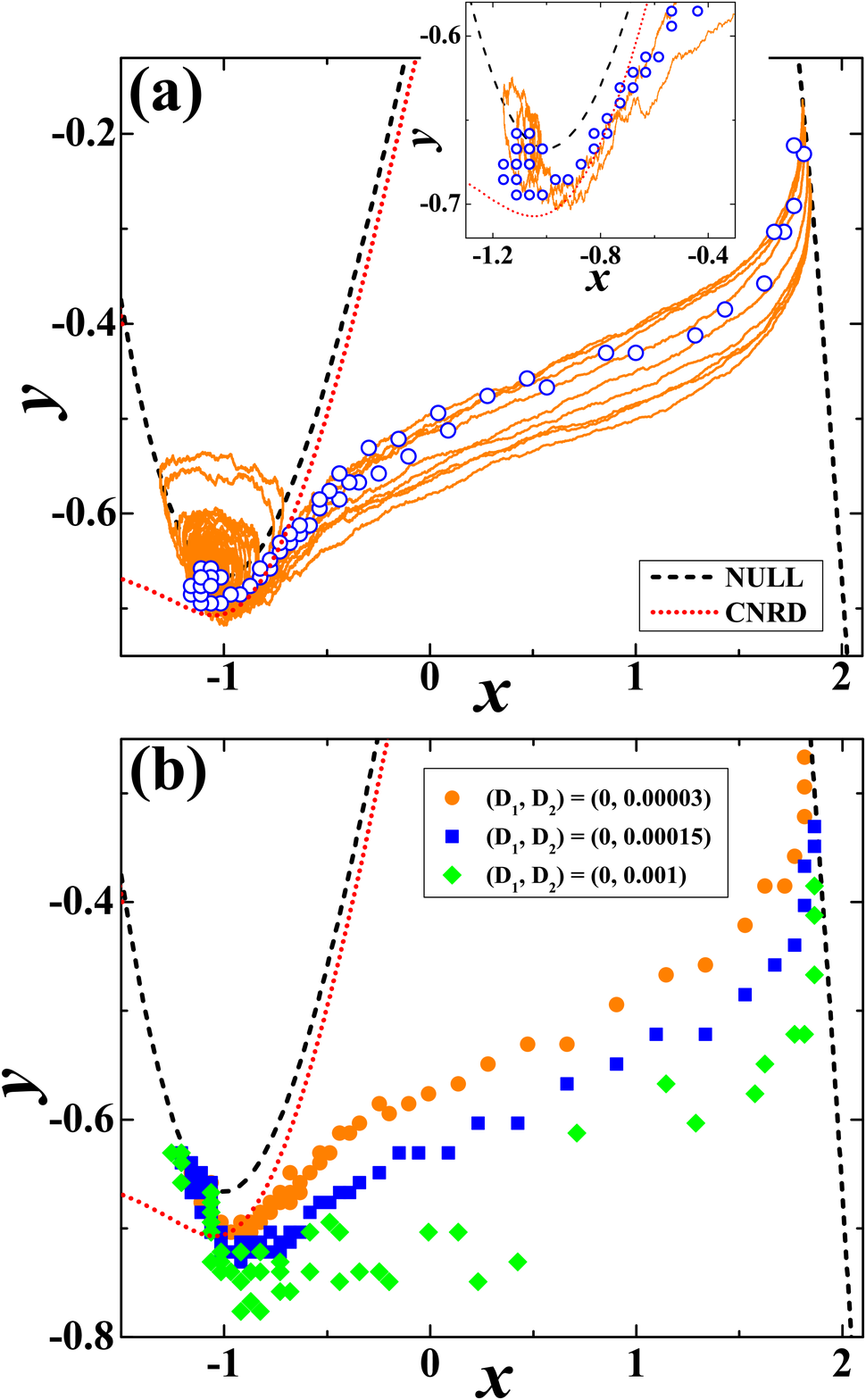,width=8.5cm}}
\caption{(Color online) $MPAP$s for an excitable $FHN$ unit. (a) The main frame shows that the activation paths for different stochastic realizations (solid lines) cluster around the $MPAP$ (open blue circles). The ratio of external vs internal noise intensities is $r_x:r_y=0:1$, whereas the noise intensity is $D=0.00003$. The $x$-nullcline $(NULL)$ and the canard-like trajectory ($CNRD$) pertaining the ghost separatrix are shown by the dashed and the dotted line, respectively. In the inset is displayed the enlarged view of the region of phase space before the $CNRD$. In (b) is illustrated how the $MPAP$ profile changes under increasing $D$ for the fixed ratio $r_x:r_y=0:1$. The $MPAP$s shown correspond to $D=0.00003$ (circles), $D=0.00015$ (squares) and $D=0.001$ (diamonds). These noise intensities lie substantially below, in vicinity and above the stochastic bifurcation, respectively.}
\label{Fig2}
\end{figure}

Let us now consider some examples of $MPAP$s in order to corroborate the points stated above. First we illustrate the validity of the method used to obtain the $MPAP$s, see Fig. \ref{Fig2}(a). The figure refers to the case $(r_x,r_y)=(0,1)$ for $D=0.00003$, the noise value substantially below the stochastic bifurcation. By taking ten arbitrary realizations of the first pulse emission process, it is shown that the stochastic realizations indeed cluster around the $MPAP$, indicated by the open circles. We have observed that the distribution of $(x,y)$ values at fixed $t$ for different stochastic realization is narrow around the terminating boundary, and broadens towards the initiation point. With increasing noise intensity $D$, the changes in the shape of the $MPAP$s become clearly visible around $D\approx0.00015$, the value close to the onset of stochastic bifurcation, cf. Fig. \ref{Fig2}(b). The effect of noise is felt particularly strong in the region of phase space before the ghost- separatrix, which is indicated by the dotted line. An interesting point is that the larger noise may also have an inhibitory effect on the process of first pulse emission, in a sense that the phase point which has already crossed the ghost separatrix may still diffuse back to the basin of attraction of the fixed point.

\begin{figure}
\centerline{\epsfig{file=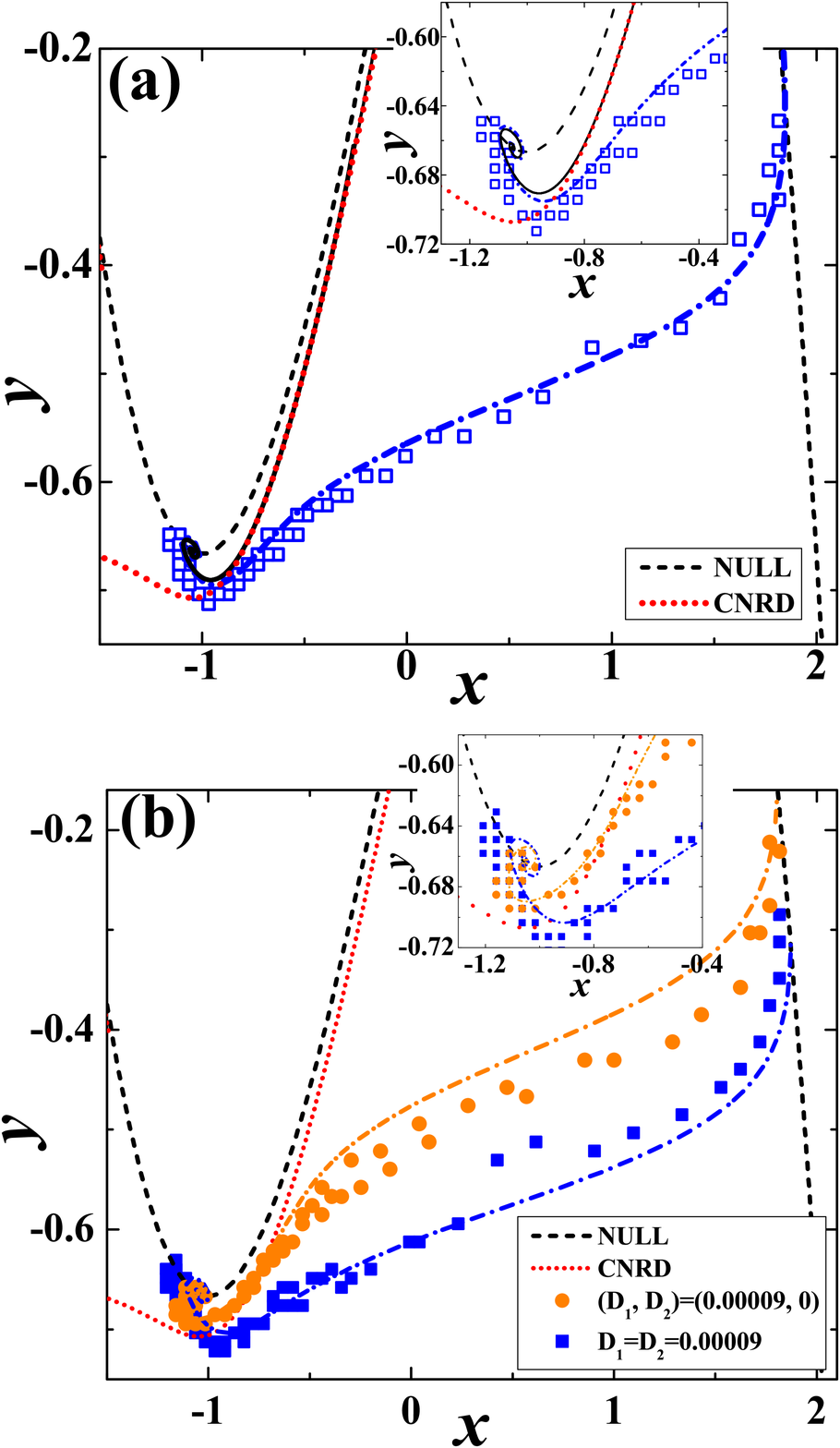,width=8.5cm}}
\caption{(Color online) Extended analysis of $MPAP$s. (a) is intended to illustrate the difference between the first pulse emission problem and the generalized escape problem for a $FHN$ unit. The setup and the style of presentation are the same as in Fig. \ref{Fig2}(a), but two additional curves are presented. The solid black line approaching the $CNRD$ indicates the optimal trajectory obtained for the escape problem via the standard Hamiltonian approach, whereas the dash-dotted line denotes the trajectory generated by the system \eqref{eq6} according to the recipe described in subsection \ref{Method}. (b) shows the  $MPAP$s for $r_x:r_y=1:0$ (circles) and  $r_x:r_y=1:1$ (squares), together with the corresponding trajectories (dash-dotted lines) generated by the system \eqref{eq6}. In both instances, the noise intensity is $D=0.00009$.}
\label{Fig3}
\end{figure}

We further highlight how the topological properties of $MPAP$s depend on the characteristic ratio $r_x/r_y$. As announced earlier, the discussion is put into a broader perspective by comparing the $MPAP$s to the trajectories generated by the effective Hamiltonian system \eqref{eq6} according to the prescription provided in Subsection \ref{Method}.  Taking the case $r_x:r_y=0:1$ as an example, we first demonstrate how the problem of first pulse emission is different from the extension of the escape problem to an excitable $FHN$ unit addressed in \cite{KPLM13}. Apart from the ghost separatrix, denoted by the dotted line, Fig. \ref{Fig3}(a) also shows the optimal trajectory for the escape problem obtained using the standard Hamiltonian formalism. The latter trajectory is indicated by the solid line which approaches, but does not cross the ghost separatrix, because the optimal escape path cannot intersect the system's manifold. The $MPAP$ numerically determined for the process of first pulse emission is represented by the open squares. The particular $MPAP$ is obtained for small, but finite noise $D$. Such $D$ values may actually be referred to as intermediate, because on one hand, they lie below the critical domain giving rise to stochastic bifurcation, but on the other hand, they cannot be considered as asymptotically small noise $(D\rightarrow 0)$ where the theory of large fluctuations would naturally apply. It can be seen that the initial part of the $MPAP$ matches very closely the optimal trajectory calculated by the standard Hamiltonian formalism, but also departs from it well before reaching the ghost separatrix. Nevertheless, an interesting finding is that the trajectory generated from the effective system \eqref{eq6}, cf. the thick solid line that crosses the ghost separatrix, matches quite closely the $MPAP$ along the entire trajectory relevant for the process of first pulse emission. This point suggests that a theory using certain aspects of the standard Hamiltonian approach could potentially be derived to characterize the first pulse emission process.

Fig. \ref{Fig3}(b) is intended to compare the $MPAP$s for the two remaining characteristic $r_x/r_y$ ratios at $D$ values substantially below the stochastic bifurcation. As expected, the respective trajectories show significant differences well before crossing the ghost separatrix. The trajectories generated by the effective Hamiltonian equations in the fashion described in Subsection \ref{Method} again seem to closely match the $MPAP$s obtained by averaging over the ensemble of stochastic realizations. The profile of the escape paths in vicinity of equilibrium in Fig. \ref{Fig3}(b) suggests that the noise-induced linearization \cite{SSSMLM96,HSE14} takes place in presence of the external noise. In particular, the latter is found to smear the effect of nonlinearity, such that the threshold-like behavior becomes smoother.

\subsection{An insight into the activation processes driven by external or internal noise} \label{Analapp}

Before proceeding to numerical results regarding the statistical properties of the activation process, let us briefly consider the conceptual differences between the cases where noise influences the dynamics on the fast ($D_1>0,D_2=0$) or the slow characteristic timescale ($D_1=0,D_2>0$). Note that the "mean activation times" obtained from approximations introduced here are not intended to be compared quantitatively with the actual stochastic averages from Sec. \ref{Numerics}, but are rather aimed at gaining qualitative insight into the distinct mechanisms by which the two noise sources affect the activation process. To this end, we use the standard approach which consists in reducing the original dynamics, given by \eqref{eq1} with $D_1$ or $D_2$ set to zero, to an appropriate Langevin equation of the form $dz=-U'(z)dt+\sqrt{2D}dW(t)$, where $U(z)$ denotes the effective potential.

If only $D_1$ is present, one may conveniently exploit the sharp separation between the two characteristic time scales. The analysis is confined solely to the fast variable subsystem, which is first rewritten as
\begin{align}
dx=-\frac{\partial U(x,y)}{\partial x}dt+\sqrt{2D_1}dW, \label{eq121}
\end{align}
with $U(x,y)=-\frac{1}{2}x^2+\frac{1}{12}x^4+xy$. In the last expression for $U$, $y$ may simply be seen as a parameter, because the evolution of the $y$-variable takes place on a timescale much slower than that of the $x$-variable \cite{DVM05}. Another useful point is that $U$ has the form of a double-well potential. Its two local minima, as well as the local maximum, correspond to $x$-values which are the roots of the equation for the $x$-nullcline $x-\frac{1}{3}x^3-y=0$. In particular, the local minima are given by the solutions $x_1$ and $x_3$ lying on the refractory and the spiking branch, respectively, while the local maximum coincides with the $x_2$ solution on the unstable branch of the $x$-nullcline ($x_1(y)<x_2(y)<x_3(y)$). Then, the activation process may be interpreted as a jump from the well at the refractory branch to the one at the spiking branch, whereby the phase point has to overcome the potential barrier provided by the local maximum.

\begin{figure}
\centerline{\epsfig{file=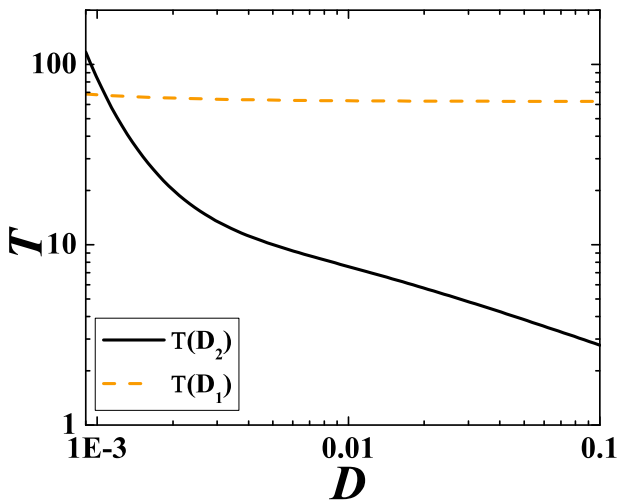,width=8.5cm}}
\caption{(Color online) Assessing the impact of two distinct noise sources on the activation process of a single
$FHN$ unit. The curves refer to approximate dependencies of "activation times" $T$ on $D_1$ (dashed line) and
$D_2$ (solid line). Note that the introduced approximations are crude, so that the results can only be considered in a qualitative fashion. Still, the two curves accurately predict that the activation process led by $D_2$ is comparably faster than the one led by $D_1$.}
\label{Fig4}
\end{figure}

Given that $U$ has a double-well shape, the escape time for a particle to jump from $x_1(y)$ to a point near $x_3(y)$ can be approximated by the well-known Kramers formula \cite{G04,R89,PB13} $T=\frac{2\pi}{\sqrt{|U"(x_2)|U"(x_1)}}e^{[(U(x_2)-U(x_1))/D_1]}$. In our case, one may fix the $y=-2/3$ value, which corresponds to the minimum of the cubic nullcline. Having determined the appropriate $x_1(-1)$ and $x_2(-1)$, we have obtained the dependence $T(D_1)$ shown by the dashed line in Fig. \ref{Fig4}. Naturally, given the crudeness of the introduced approximations, both in terms of system dynamics and the boundary conditions, the obtained $T$ values should not be compared to the mean activation times for our activation problem.

The scenario where only $D_2$ is present cannot be approached the same way as above, because $y$ can no longer be treated as a parameter. Still, one may implement the adiabatic elimination of the fast variable to find the effective potential that influences the $y$ dynamics. Though the roots of the equation describing the slow manifold $y=x-\frac{1}{3}x^3$ may be used explicitly, the drawback is that such solutions are not easily handled  analytically. In order to keep the subsequent form of effective potential analytically tractable, it is more convenient to approximate the relation between $y$ and $x$ in vicinity of the minimum of the cubic nullcline by a simple relation $y(x)=y_m+\frac{1}{2}k(x-x_m)^2$, where $(x_m,y_m)=(-1,-2/3)$ refer to coordinates of the minimum. In other words, in proximity of the minimum, the cubic dependence has been replaced by a quadratic approximation. Note that $k=-2x_m$ is determined by taking the second derivative of the equation for the $x$-nullcline. From the expression for $y(x)$, one readily finds that $x=x_m\pm\sqrt{y-y_m}$, whereby the plus sign solution is relevant for our activation problem. It follows that the equation for the dynamics of $y$ may be written as
\begin{align}
dy=\epsilon(x_m+\sqrt{y-y_m}+b)dt+\sqrt{2D_2}dW, \label{eq122}
\end{align}
such that the corresponding effective potential reads $U=\epsilon(x_m+b)y+\frac{2}{3}\epsilon(y-y_m)^{3/2}$.

Given that $U$ is not a double-well potential, one cannot use the Kramers-like equation for the mean first-exit time. Instead, it is appropriate to use the general form
\begin{align}
T=\frac{1}{D_2}\int_{i}^{a}du e^{U(u)/D_2}\int_{r}^{u}dv e^{-U(v)/D_2}, \label{eq123}
\end{align}
derived from the Fokker-Planck approximation to a problem involving a single absorbing boundary $a$ and a single reflecting boundary $r$ \cite{G04,R89,PB13}. Note that $i$ in one of the integration limits refers to the injection point (initial location), which in our activation problem corresponds to the deterministic fixed point $(x_{eq},y_{eq})=(-b,-b+b^3/3)$. Regarding the reflecting boundary, it has often been found that the final results are not affected by its particular value \cite{HE05,G04}, so $r$ may readily be set to $r\rightarrow-\infty$ to simplify the calculations. As for the absorbing boundary, it generally concerns the terminating point of the activation path and remains a free parameter that cannot be known \emph{a priori}. Nevertheless, we may use the results for the $MPAP$ solutions from Sec. \ref{Examples}, and simply read the necessary coordinates of the terminating point ($y=-0.308$), which illustrates the complementary nature of the methods applied. Note that the integrals such as the ones in \eqref{eq123} are typically resolved by introducing convenient approximations. In the particular case, we have used the approximate solution provided in \cite{HE05}. The obtained curve $T(D_2)$ is shown by the solid line in Fig. \ref{Fig4}.

It has already explained that the results in Fig. \ref{Fig4} cannot be considered reliable in quantitative sense given the crudeness of the approximations involved. Still, certain qualitative insight have been gained. For instance, the analysis above indicates that the activation processes led by $D_1$ or $D_2$ have two considerably distinct backgrounds, whereby only the mechanism of the former may be interpreted in analogy to a jump over the barrier that separates two potential wells. Further, the curves in Fig. \ref{Fig4} turn out to be consistent with the general trend for a single $FHN$ unit demonstrated in the next subsection, according to which the activation is more easily excited by $D_2$ than $D_1$.

\subsection{Statistical features of the activation process} \label{Numerics}

\begin{figure}
\centerline{\epsfig{file=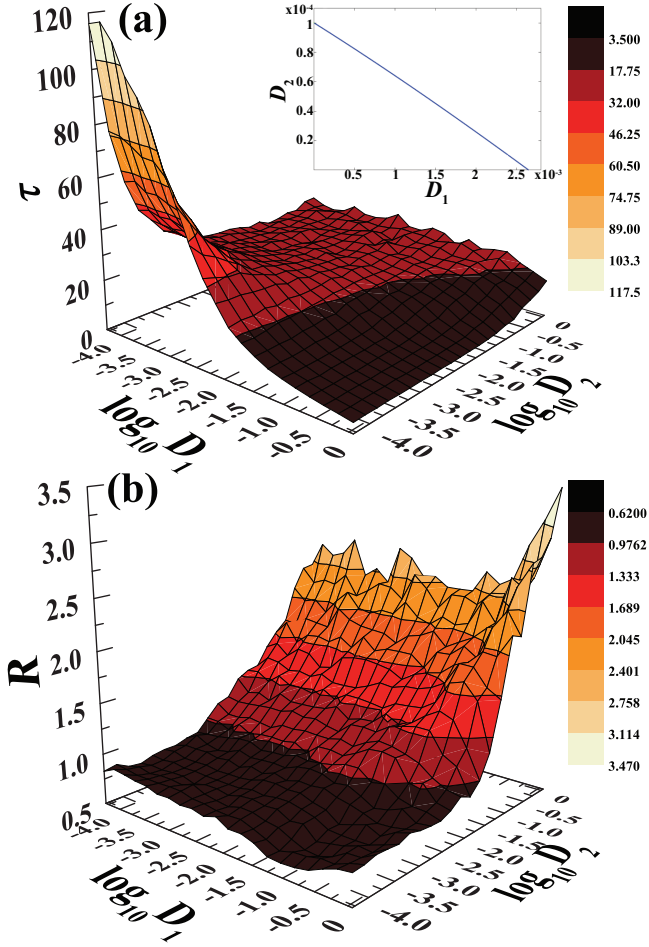,width=8.5cm}}
\caption{(Color online) Statistical features of activation process influenced by $D_1$ and $D_2$. In (a) is displayed how the mean $TFP$s $\tau$ averaged over an ensemble of different stochastic realizations depend on $D_1$ and $D_2$, whereas (b) concerns the associated coefficient of variation $R(D_1,D_2)$. $\tau(D_1,D_2)$ exhibits three characteristic regimes of behavior, whereby the transition from the large values to the plateau region is found to qualitatively correspond to stochastic bifurcation from the stochastically stable fixed point to continuous oscillations. An indication on the noise intensities that give rise to stochastic bifurcation is provided in the inset of (a). The latter shows bifurcation curve $D_2(D_1)$ obtained for the approximate model \eqref{eq13} of the original stochastic system \eqref{eq1}.}
\label{Fig5}
\end{figure}

The statistics of activation events is characterized in terms of the mean $TFP$ $\tau(D_1,D_2)$ and the associated coefficient of variation $R(D_1,D_2)$. The former is an average of $TFP$s for different stochastic realizations $\tau(D_1,D_2)=\frac{1}{n_r}\sum\limits_{i=1}^{n_r}\tau_i(D_1,D_2)$. The stochastic paths taken into account satisfy the specified boundary conditions, such that the trajectories emanate from the deterministic fixed point and terminate at the spiking branch of the cubic nullcline. The coefficient of variation is defined as the normalized variation of activation times \cite{PK97}
\begin{align}
R(D_1,D_2)=\frac{\sqrt{\langle \tau_i^2\rangle-\langle\tau_i\rangle^2}}{\langle \tau_i\rangle},  \label{eq12}
\end{align}
where $\langle \cdot\rangle$ refers to averaging over an ensemble of stochastic realizations. Quantity $R$ is intended to describe the regularity of the activation process, in a sense that the smaller $R$ indicates that the $TFP$s deviate less from the mean value. Note that numerical simulations for all the considered setups are carried out by implementing the Euler integration scheme with the fixed time step $\delta t= 0.002$, having verified that no changes occur for smaller $\delta t$. All the results for the mean $TFP$s and the associated variances are obtained by averaging over an ensemble of $5000$ different stochastic realizations of the activation process. In each realization, the initial conditions of a unit coincide with the deterministic fixed point. 

The fields $\tau(D_1,D_2)$ and $R(D_1,D_2)$ for a single excitable unit are plotted in Fig. \ref{Fig5}(a) and Fig. \ref{Fig5}(b), respectively. The obtained profiles indicate that the mean $TFP$s are more sensitive to variation of $D_1$, whereas $R$ shows strong dependence on $D_2$. Regarding the mean $TFP$s, one may distinguish three characteristic regimes, including $(i)$ long $TFP$s, encountered for small $D_1$ and $D_2$, $(ii)$ the plateau region, comprising intermediate $D_1$ and intermediate to large $D_2$ values, as well as $(iii)$ short $TFP$s,
found for large $D_1$ irrespective of $D_2$.

Given that the considered stochastic process is influenced by two sources of noise, one finds that the transitions between the different regimes are gradual rather than sharp, whereby the "boundaries" are naturally smeared due to action of noise. Nevertheless, in terms of theory, it is reasonable and well justified to discuss the physical background giving rise to transitions between the different regimes. What we postulate is that the transition between the domains $(i)$ and $(ii)$ may qualitatively be accounted for by the fact that the excitable unit undergoes stochastic bifurcation induced by $D_1$ and $D_2$. Note that the phenomenological stochastic bifurcation \cite{A99, ABR04, GLV09, GBV11} we refer to corresponds to the noise-induced transition from the stochastically stable fixed point (stationary probability distribution $P(x,y)$ focused around the fixed point) to the stochastically stable limit cycle (stationary probability distribution $P(x,y)$ showing non-negligible contribution for $(x,y)$ values along the spiking and the refractory branches of the $x$-nullcline). Intuitively, one understands that the fixed point can be considered stochastically stable if the amplitude of fluctuations around the fixed point is of the order of noise intensity. By the same token, one may perceive a limit cycle as stochastically stable if the general structure involving two branches of the $x$-nullcline (the spiking and the refractory branch) is preserved under the action of noise.

In support of associating the transition between the domains of large $TFP$s and the plateau region with the stochastic bifurcation, one may invoke the qualitative argument that above the bifurcation, the attractive power of the fixed point is effectively no longer felt, in a sense that noise can drive the phase point away from equilibrium without meeting a strong opposing force. At the level of mean $TFP$s, this should be reflected as follows. Below the stochastic bifurcation, the mean $TFP$s are expected to be longer, while above the bifurcation they should substantially reduce and also become fairly insensitive to further increase of noise. In other words, once the fixed point becomes stochastically unstable, the gross effect of noise is the same because the fixed point holds no attractive power to resist its action.

Having established that the boundary between the domain of long $TFP$s and the plateau region in Fig. \ref{Fig5}(a) should coincide with the $(D_1,D_2)$ values that give rise to stochastic bifurcation, our next goal is to try to obtain these values analytically. To do so, we derive a deterministic model based on Gaussian approximation of the original system \eqref{eq1}. According to Gaussian approximation, all the cumulants above the second order are assumed to vanish \cite{TP01,FTVB13,BRTV10}. One is ultimately left with a set of five equations describing the dynamics of the first moments $m_x(t)=E[x(t)]$ and $m_y=E[y(t)]$, the variances $s_x(t)=E[(x(t)-m_x(t))^2]$ and $s_y(t)=E[(y(t)-m_y(t))^2]$, as well as the covariance $u(t)=E[(x(t)-m_x(t))(y(t)-m_y(t))]$:
\begin{align}
\dot{m_x}&=m_x-\frac{1}{3}m_x^3-m_xs_x-m_y \nonumber\\
\dot{m_y}&=\epsilon(m_x+b)\nonumber\\
\dot{s_x}&=2s_x(1-m_x^2-s_x)-2u+2D_1 \nonumber\\
\dot{s_y}&=2\epsilon u+2D_2 \nonumber\\
\dot{u}&=u(1-m_x^2-s_x)+\epsilon s_x-s_y. \label{eq13}
\end{align}
Note that the impact of noise in \eqref{eq13} is described only by the respective intensities $D_1$ and $D_2$, which can then be treated as bifurcation parameters. In particular, the bifurcation analysis shows that the approximate model \eqref{eq13} displays direct supercritical Hopf bifurcation, which qualitatively accounts for the stochastic bifurcation exhibited by the excitable unit \eqref{eq1}. The obtained bifurcation curve $D_2(D_1)$, plotted in the inset of Fig. \ref{Fig5}(a), and hence the transition from the large values of mean $TFP$s to the plateau region.

Better understanding of the nature of the activation process behind the $\tau$ and $R$ dependencies from Fig. \ref{Fig5}(a) and Fig. \ref{Fig5}(b) may be gained by examining how the distribution of $TFP$s over an ensemble of different stochastic realizations changes under variation of $D_1$ and $D_2$. Let us note first that even for a typical escape problem, such an issue has rarely been addressed in the literature and is difficult to approach analytically. In particular, for the distribution of first exit times in a typical escape problem, the only rigorous result so far has been found within the framework of large fluctuations theory \cite{D83}. It states that for the \emph{sufficiently small} noise intensity, the distribution of the first exit times \emph{asymptotically} acquires exponential form. If interpolated to the case of an excitable $FHN$ unit, one should expect the latter statement to apply regardless of whether noise is added solely to the fast or the slow variable. Still, note that the mentioned result could concern only the extension of the escape problem to excitable systems, where the terminating boundary set would be given by the "ghost separatrix". Recall that we have already explained the conceptual difference between such a problem and the activation problem we consider, where the focus exclusively lies with the spiking response of an excitable unit.

\begin{figure}
\centerline{\epsfig{file=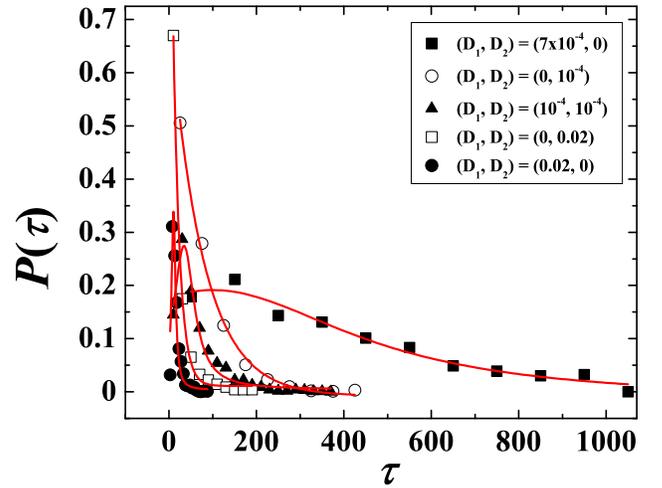,width=8.5cm}}
\caption{(Color online) Impact of $D_1$ and $D_2$ on the distribution of $TFP$s $P(\tau)$ obtained for an ensemble of different stochastic realizations. $D_2$ alone typically gives rise to an exponential distribution of $TFP$s, which is consistent with the Poisson process. Under prevailing $D_1$, the distributions are found to conform to a unimodal, Lorentzian-like profile. The displayed examples refer to cases $D_1=0.0007,D_2=0$ (solid squares),
$D_1=0.0001,D_2=0.0001$ (solid triangles), $D_1=0.02,D_2=0$ (solid circles), $D_1=0,D_2=0.0001$ (empty circles) and $D_1=0,D_2=0.02$ (empty squares).}
\label{Fig6}
\end{figure}

The characteristic examples illustrated in Fig. \ref{Fig6} suggest that the activation process dominated by $D_2$ gives rise to the exponential distribution of $TFP$s, whereas the distributions generated by the prevailing $D_1$ conform to Lorentzian-like profile with the cut-off at small $TFP$ values. Regarding the former, one should emphasize that the exponential distribution of the inter-event intervals is typically associated with the Poisson process \cite{DA01}, and it is indeed not unexpected to find an excitable unit under small amount of noise to act as a Poisson generator \cite{HE05}. Nevertheless, an interesting finding is that the activation events led by $D_1$ seem to be derived from some process other than the Poissonian, though the asymptotic distribution of $TFP$s is still exponential. Note that the qualitative distinction between the average effects of $D_1$ and $D_2$ has already been commented on in subsection \ref{Single}. The remark on the two distribution types holds if the leading noise term is much stronger than the remaining one. Nevertheless, while the validity of this statement is maintained even under substantial increase of $D_1$ as the leading term, the analogous point for $D_2$ applies up to a certain value, as explained below.

For $(D_1,D_2)$ corresponding to the plateau from Fig. \ref{Fig5}(a), one generally encounters the exponential distribution of $TFP$s or some of its modifications. In this context, note that the increase of $R$ in Fig. \ref{Fig5}(b) remains fairly slow, if any, for $D_2$ values that warrant $R<1$, but becomes steep once $R$ passes $1$ around $D_2\sim10^{-2}$. The approximate boundary at $R=1$ coincides with the coefficient of variation one obtains for the exponential distribution of $TFP$s. It is further found that the sufficiently large $D_2$ values where $R>1$ give rise to a peculiar regime where the ensemble of activation events splits in two sharply distinct classes, such that the one with the small $TFP$s dominates, but the contribution from the events with large $TFP$s is non-negligible. The examples of stochastic activation paths for large $D_2$, including those with long $TFP$s, where the phase point rebounds onto the refractory branch of the $x$-nullcline before the transition to spiking branch is triggered, are already provided in Fig. \ref{Fig3}(b). Note that the intention there has been to illustrate the setup for which the analytical method of calculating the $MPAP$s fails.

\section{Case of two units: $MPAP$s and statistical properties of the activation process}  \label{Twounits}

In this Section, we investigate how the form of coupling affects the first pulse emission process for two units subjected to external and internal noise. The first subsection is focused on the $MPAP$s obtained for the setups with linear or nonlinear interactions, whereas the second subsection concerns the statistical features of the activation process.

\subsection{Dependence of $MPAP$s on the form of coupling}
\label{Twopaths}

The $MPAP$s are obtained by the numerical method presented in subsection \ref{Method}. Nevertheless, before proceeding with the results, we first consider what counts as a two-unit activation event by providing the appropriate boundary conditions. In terms of the initial conditions, the activation paths of units described by \eqref{eq2} or \eqref{eq3} begin at the equilibrium
$(x_{1,eq},y_{1,eq},x_{2,eq},y_{2,eq})=(-b,-b+b^3/3,-b,-b+b^3/3)$. The terminating boundary conditions are specified in such a way that the phase points of \emph{both} units should reach the spiking branch of their respective $x_i$-nullclines. This definition implies that the time-to-first pulse for a couple of units is determined by the slower-firing unit. Note that the profile of the spiking branch for a coupled unit is quite similar to that of a single unit, which can readily be verified. In particular, for the given unit one may approximate the $x$-coordinate of the other unit within the coupling term by its value at the deterministic fixed point, around which the actual values fluctuate for most of the time.

\begin{figure}
\centerline{\epsfig{file=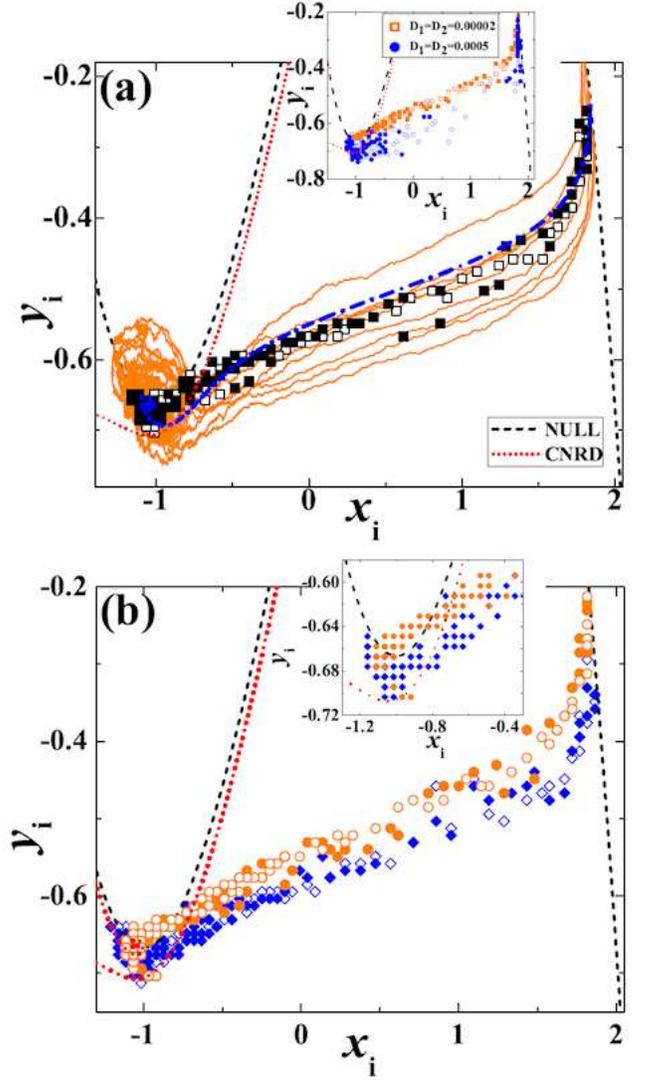,width=8.5cm}}
\caption{(Color online) $MPAP$s for two coupled units. In (a) are displayed the $MPAP$s for two units (open and solid squares) interacting via linear couplings of subcritical strength $c=0.04$. The solid lines denote ten sample paths corresponding to the process of first pulse emission. The results are obtained for $r_x:r_y=1:1$ and $D=0.00005$. The dash-dotted line indicates the trajectory generated by the extended system \eqref{eq14}. In the inset are shown the $MPAP$s for the same characteristic ratio and noise intensities $D=0.00002$ (squares) and $D=0.0005$ (circles). (b) provides a comparison between the $MPAP$s obtained for linear (open and solid circles) and nonlinear interactions (open and solid diamonds). In the latter case, the subcritical coupling strength is $c=0.07$, while the noise parameters are the same as in (a). In the inset is shown the enlarged view of the portion of phase space before the $CNRD$.}
\label{Fig7}
\end{figure}

Regarding the coupling strengths, note that we are interested only in $c$ values that are subcritical with respect to Hopf bifurcation, viz. $c$ is always set so that the deterministic versions of \eqref{eq2} or \eqref{eq3} admit the locally stable fixed point. This is consistent with the goal to examine the \emph{noise-driven} first pulse emission process and the fashion in which it is modified by the presence of coupling. The fact that the $c$ values are subcritical also implies that in the deterministic case, the large excitation of one unit (initial conditions far from equilibrium) cannot induce a spiking response of the other unit which lies at equilibrium.
One should point out that the adopted formulation of the activation problem for two units is by far more appropriate than the alternatives, including the one cast in terms of the average variables $(x_1+x_2)/2$ and $(y_1+y_2)/2$. Without stating the details, we make a remark that the latter approach would have several drawbacks. On one hand, the dynamics associated to synchronization would smear the physical picture relevant for the activation problem, whereas on the other hand, there would be no immediate generalization from the case of a single unit to the setups with two units.

Now let us focus on the scenario where two excitable units interact via linear couplings. The symmetry of interactions is reflected in the point that the $MPAP$ trajectories are identical for both units. This is corroborated in Fig. \ref{Fig7}(a), which shows as an example the respective $MPAP$s of the two units (solid and open squares) for the characteristic ratio $r_x:r_y=1:1$ at small $D$ substantially below the stochastic bifurcation. Several realizations of the first pulse emission process are plotted to demonstrate the clustering around the $MPAP$. The approximate matching between the units' most likely trajectories is found to persist for all $(r_x,r_y)$ characteristic setups. The changes of the $MPAP$ profile under increasing noise intensity $D$ are illustrated in the inset of Fig. \ref{Fig7}(a). It has already been mentioned that depending on $c$, the units coupled in a linear fashion may display two different regimes where the fixed point is stable. These regimes have been referred to as excitability proper and generalized excitability, whereby the latter involves coexistence between the stable fixed point and the stable limit cycle created in a global fold-cycle bifurcation. Comparing the cases where $c$ is sub- or supercritical with respect to global bifurcation, we have established that for small but finite $D$, the profile of $MPAP$s for all three characteristic ratios $r_x/r_y$ does not appear to show significant differences.

As in case of a single unit, we make a brief remark regarding the trajectories generated by the effective set of Hamiltonian equations. Compared to \eqref{eq6}, the equations for the dynamics of the extended system are modified to include the interaction terms
\begin{align}
\frac{dx_i}{dt}&=x_i-x_i^3/3-y_i+r_xp_{x,i}+c(x_i-x_j) \nonumber\\
\frac{dy_i}{dt}&=\epsilon(x_i+b)+r_yp_{y,i} \nonumber\\
\frac{dp_{x,i}}{dt}&=-(1-x_i^2)p_{x,i}-\epsilon p_{y,i}+cp_{x,j}\nonumber\\
\frac{dp_{y,i}}{dt}&=p_{x,i}, \label{eq14}
\end{align}
whereby $i,j\in\{1,2\},i\neq j$ denote the units' indices. The numerical treatment of the above system again involves integration for the initial conditions set on the unstable manifold of the saddle point. In particular, in configuration subspace one chooses the initial values $(x_i,y_i)$ that lie on a four-dimensional sphere of a very small radius, which encloses the deterministic fixed point $(x_{1,eq},y_{1,eq},x_{2,eq},y_{2,eq})$. Naturally, the points located on a four dimensional sphere are parametrized with three independent angular variables. The initial values of the generalized momenta are then obtained using the prescription provided in \cite{BMLSM05}. An interesting finding is that the trajectories of the extended system, selected by the rule described in subsection \ref{Method}, again closely match the numerically obtained $MPAP$s, cf. the dotted line in Fig. \ref{Fig7}(a). This striking similarity further evinces that the theory in the spirit of the Hamiltonian approach may potentially be derived for the problem of first pulse emission.

An issue that requires to be addressed is whether and how sensitive are the topological features of the $MPAP$s with respect to linear/nonlinear form of coupling. The comparison is facilitated in Fig. \ref{Fig7}(b), where the corresponding $MPAP$s for the scenarios with linear (circles) and nonlinear interactions (diamonds) are plotted together. The data shown are obtained for the same $D_1$ and $D_2$, whereas the coupling strengths are analogous in terms of distance from the Hopf bifurcation. The trajectories corresponding to the different units for the same system configuration are distinguished by the solid and open symbols. Note that in case of nonlinear interactions, there is no symmetry to warrant that the respective $MPAP$s of the two units would be identical for arbitrary values of external and internal noise. It turns out that the $MPAP$s are approximately identical for intermediate $D$ sufficiently below the stochastic bifurcation. Nevertheless, we have also found that the spread of the $MPAP$s in case of nonlinear couplings increases with $D$. As for the effects of linear vs nonlinear interactions, one observes significant differences in the profiles of $MPAP$s within the initiation region, which is shown enlarged in the inset of Fig. \ref{Fig7}(b). Also note that the $MPAP$s of the interacting units substantially depart from what is found in case of a single unit, cf. Fig. \ref{Fig2} and Fig. \ref{Fig3}.

\subsection{Statistical properties of the activation process for coupled units}
\label{StatCouple}

\begin{figure*}
\centerline{\epsfig{file=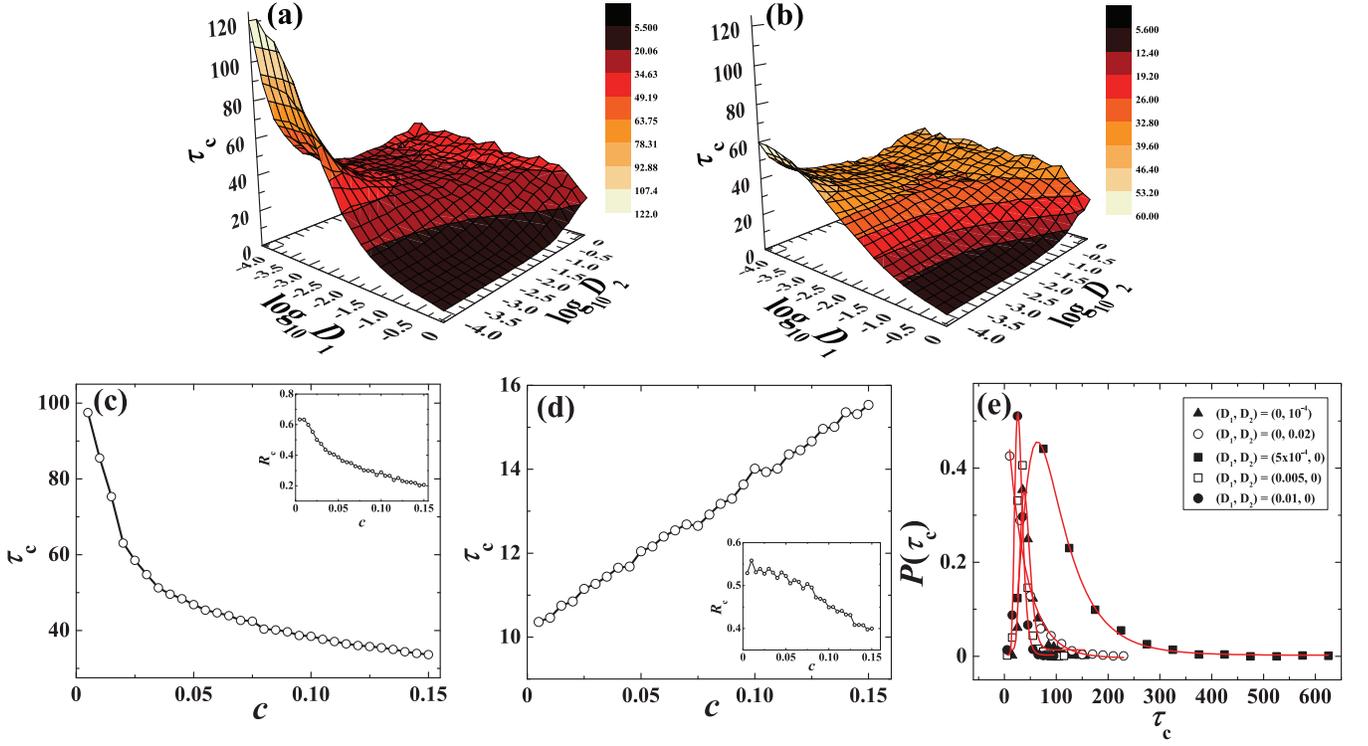,width=18cm}}
\caption{(Color online) Statistical properties of activation process for two units interacting via linear couplings. (a) and (b) refer to mean $TFP$s $\tau_c(D_1,D_2)$ in a strongly subcritical ($c=0.01$) and a weakly subcritical regime $(c=0.04)$, respectively. Note that the definition of the two-unit activation event requires that the phase points of both units have reached the appropriate terminating boundary set. (c) and (d) show how $\tau_c$ (main frames) and $R_c$ (insets) behave under increasing $c$. The noise intensities $(D_1,D_2)=(0.00014,0.0008)$ fixed in (c) are representative for the domain of large $TFP$s from (a), whereas (d) is obtained for $(D_1,D_2)=(0.154,0.0002)$, the values corresponding to the plateau region in (a). In (e) is demonstrated how the distributions of $TFP$s over different stochastic realizations vary with $D_1$ and $D_2$. Note that the unimodal distribution profile is preferred over the exponential form. The data are obtained for $D_1=0,D_2=0.0001$ (solid triangles), $D_1=0.0005,D_2=0$ (solid squares), $D_1=0.01,D_2=0$ (solid circles), $D_1=0,D_2=0.02$ (empty circles) and $D_1=0.005,D_2=0$ (empty squares).}
\label{Fig8}
\end{figure*}

Here we consider two types of numerical results, one referring to the two-unit activation events, and the other concerning the correlation between the activation events on individual units. Recall that the "compound" activation event for a couple of units requires that the phase points of both units have reached the spiking branch of the appropriate $x_i$-nullcline, consistent with the definition adopted in Sec. \ref{Twopaths}. Adhering to this, we
have calculated the mean $TFP$s $\tau_c(D_1,D_2)$ and the associated coefficients of variation $R_c(D_1,D_2)$ for the scenarios with the linear or the nonlinear couplings, further examining how the results are changed under variation of $c$. Though $c$ is always selected to lie below the Hopf threshold, in the latter context one may still distinguish between the strongly and the weakly subcritical regimes.

In terms of whether the mere form of coupling affects the statistical features of the activation process, it may be shown that the fields $\tau_c$ and $R_c$ exhibit qualitatively analogous dependencies for the linear and the nonlinear interactions. It is further found that the behavior of mean $TFP$s is in several aspects different to that of a single unit, whereas the corresponding coefficient of variation $R_c$ is only marginally dependent on $c$, displaying the "universal" behavior qualitatively similar to the one in Fig. \ref{Fig5}(b). As an example on how the properties of activation process change with $c$, in Fig. \ref{Fig8}(a) and Fig. \ref{Fig8}(b) are illustrated the respective fields $\tau_c(D_1,D_2)$ for the strongly and weakly subcritical regimes in case of the linear coupling. Both plots corroborate the existence of the three typical regimes already indicated in reference to Fig. \ref{Fig5}(a), though one no longer speaks of stochastic bifurcation in vicinity of the Hopf bifurcation controlled by $b$, but rather of the one induced by the coupling strength.

Nevertheless, several differences due to presence of coupling should be noted. First, as $c$ is increased, the mean $TFP$s for small noise intensities (below the stochastic bifurcation) become shorter, see Fig. \ref{Fig8}(c). Also, at small noise intensities the activation times $\tau_c$ of the couple are reduced when compared to the case of a single unit. However, for intermediate $D_1$ and $D_2$ corresponding to the plateau region, the mean $TFP$s for the pair are larger than those found for a single unit. More importantly, the average activation times in this region seem to increase as $c$ rises, see Fig. \ref{Fig8}(d). One may in fact discern a general trend that the differences between the three characteristic regions are gradually washed out as $c$ is enhanced, which is manifested even more once $c$ is supercritical. This is to be expected in the latter case, given that the activation process then becomes primarily deterministic, viz. it is only perturbed by the noise terms. Note that in all the three characteristic regimes the corresponding coefficients of variation are seen to reduce with $c$, cf. the insets in Fig. \ref{Fig8}(c) and Fig. \ref{Fig8}(d). The final remark on the impact of coupling is that the region with small mean $TFP$s apparently decreases with $c$.

We have further examined how the distribution of activation events over different stochastic realizations depends on the form and strength of coupling. One may state that the general conclusions reached in case of a single unit, cf. Fig. \ref{Fig6}, persist for the activation events of coupled units, though the physical picture is more perturbed for the scenario with the linear than the nonlinear interactions. Recall that for a single unit, only the external noise has been found to induce deviations from the typical exponential distribution of events. For the scenario involving the nonlinear coupling, there may be more noise domains admitting some distribution other than exponential, but the latter still constitutes the prevailing form of behavior. However, under linear interactions, it turns out that both $D_1$ and $D_2$ may give rise to a form of $TFP$ distribution completely absent in case of a single unit. The particular form is unimodal, and is numerically best approximated by the lognormal profile, cf. Fig. \ref{Fig8}(e). One cannot suspect on the type of activation process producing such a distribution, though the explanation on why it is different from both the scenarios with a single unit and two units interacting in a nonlinear fashion likely has to take into account the existence of global bifurcation \cite{LL10}.

As far as the effects of coupling strength are concerned, the $TFP$ distributions typically become narrower as $c$ approaches the critical value, viz. the tail at longer activation times is reduced compared to an uncoupled unit for the same $(D_1,D_2)$. This point holds independent on the linear/nonlinear form of coupling. Thus, one may state that the impact of stronger coupling is expectedly reflected in the decrease of fraction of the longer individual activation events.

\begin{figure*}
\centerline{\epsfig{file=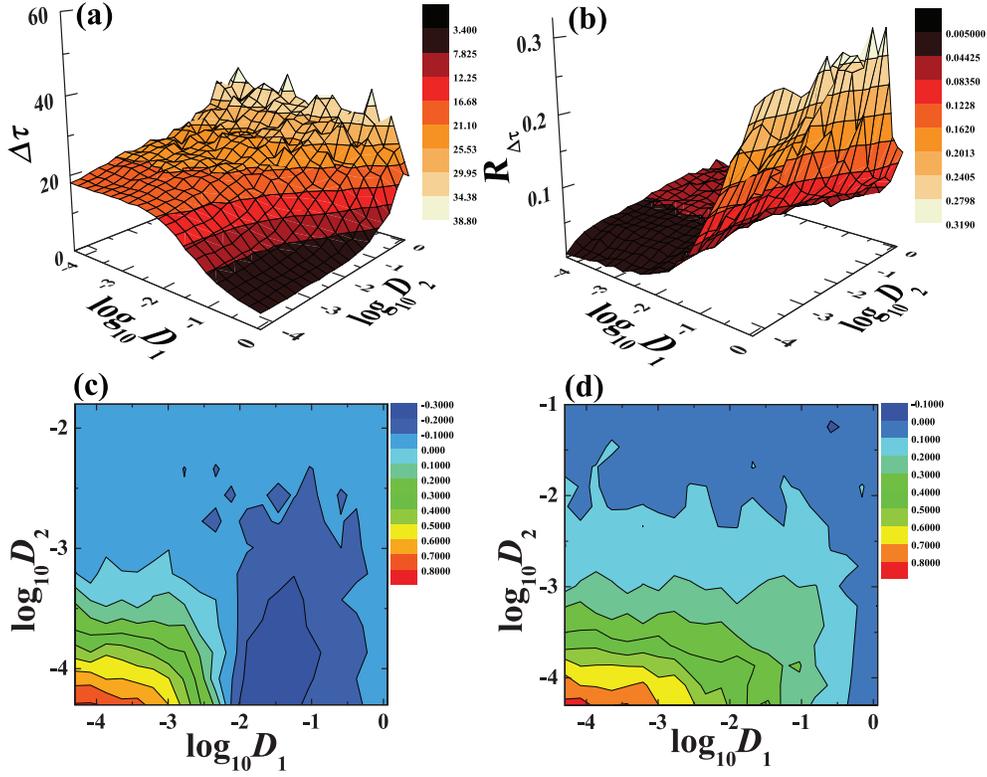,width=13cm}}
\caption{(Color online) Relationship between $TFP$s of individual units making up the pair. In (a) and (b) are plotted the differences in mean $TFP$s $\Delta \tau(D_1,D_2)$ and the associated coefficients of variation $R_{\Delta \tau}(D_1,D_2)$, respectively. The data are obtained for the linear couplings of strength $c=0.04$, and
similar results are found for the setup with nonlinear interactions. (c) and (d) show the correlation coefficients $\rho(D_1,D_2)$ between the $TFP$s for the different stochastic realizations in cases of linear $(c=0.04)$ and nonlinear interactions $(c=0.06)$, respectively.} 
\label{Fig9}
\end{figure*}

The final point we address concerns the relation between the individual activation processes on the units making up the pair. This issue may be approached from two angles, either by examining the mean difference in the single unit $TFP$s, or by analyzing the correlation between the individual activation events. On the former point, we introduce a measure of coherence between the individual activation events averaged over an ensemble of $n_r$ different stochastic realizations, $\Delta\tau=\frac{1}{n_r}\sum\limits_{k=1}^{n_r}|\tau_{k,1}-\tau_{k,2}|$, as well as the associated coefficient of variation $R_{\Delta \tau}$. $\Delta\tau(D_1,D_2)$ is intended to describe the net effect of how much the interaction is able to enforce matching between the activation times of noise-driven units. To understand the relevance of this, one should recall that the coherence of the first units' responses is an issue quite distinct from that of asymptotic synchronization between the spiking series.

In Fig. \ref{Fig9}(a) and Fig. \ref{Fig9}(b) are plotted the dependencies $\Delta \tau(D_1,D_2)$ and
$R_{\Delta \tau}(D_1,D_2)$ for the case of two units interacting via linear couplings, but qualitatively similar results are found for the nonlinear interactions as well. From Fig. \ref{Fig9}(a) one learns of a general tendency for the mean difference to increase with the internal noise. For smaller fixed $D_2$, the spread of individual $TFP$s reduces under the action of external noise, but such an effect is lost once $D_2$ overwhelms the system dynamics. Note that for large internal noise $\Delta \tau$ becomes comparable to $\tau_c$ for the coupled units, which suggests a typical scenario where the activation process of one unit is rapid, whereas the pulse emission of the other unit is substantially delayed. Naturally, for the $(D_1,D_2)$ domain supporting small $\tau_c$, the $\Delta \tau$ dependence tells us that both the units emit pulses virtually at the same time. Regarding Fig. \ref{Fig9}(b), the gross effect is that the variation of the difference of single unit $TFP$s shows a steep increase above some "threshold" external noise, whose value becomes larger as $D_2$ is enhanced. Similar dependence is found for the nonlinear interactions, but with the reverse roles of $D_1$ and $D_2$.

Next we examine the correlation of individual activation times for units comprising the pair. The correlation is quantified by the correlation coefficient between the times-to-first-pulses of single units
$\rho=\frac{\langle\tau_{k,1}\tau_{k,2}\rangle-\langle\tau_{k,1}\rangle\langle\tau_{k,2}\rangle}{\sqrt{\langle\tau_{k,1}^2\rangle-\langle\tau_{k,1}\rangle^2}\sqrt{\langle\tau_{k,2}^2\rangle-\langle\tau_{k,2}\rangle^2}}$, where the angled brackets denote averaging over an ensemble of different stochastic realizations. The fields $\rho(D_1,D_2)$ for the setups with the linear and nonlinear interactions are shown in Fig. \ref{Fig9}(c) and Fig. \ref{Fig9}(d), respectively. A common ingredient in both cases is that there exists a certain value of $D_2$, above which the noise effects prevail. There the first-time pulses are neither correlated nor anti-correlated, viz. the correlation coefficient lies around zero. However, for smaller $D_2$, the correlation between the individual activation events substantially depends on the form of coupling. For linear interactions, small $D_1$ then facilitates strong correlation, whereas for larger $D_1$ the $TFP$s become significantly anti-correlated. On the other hand, for the nonlinear couplings $\rho$ displays a more homogeneous dependence on $D_1$ and $D_2$. In fact, the correlation is strong for small noise intensities, and it reduces with the increase of both $D_1$ and $D_2$, whereby the rate of decline depends more on $D_2$ than $D_1$.

As an interesting remark for future study, we indicate a potential link between the form of correlation of the
single unit $TFP$s and the synchronization state of the stochastic units' time series
\cite{BJPS09,HYPS99,ZKH03,N94,HZ01,HL03,P05,P07} under the given parameter set.
In particular, note than in case of linear interactions the time series of two units display a constant phase shift for small $D_1$, which coincides with the domain where the $TFP$s are correlated. Nonetheless, the time series show phase slips and amplitude fluctuations for larger $D_1$, where the activation times are uncorrelated or anti-correlated. In a similar fashion, for the nonlinear coupling one typically encounters in-phase synchronization, at least for sufficiently small $D_1$, precisely where the $TFP$s are correlated. These arguments suggest that the approximately constant phase shift in the ensuing time series should imply a strong correlation between the $TFP$s of individual units. In other words, the ordering effect of coupling can make an impact on the activation processes of units in a fashion similar to what is found for long time asymptotic processes, such as synchronization between the unit's time series.

\section{Conclusions}
\label{Conc}

We have analyzed the noise-driven first-pulse emission process for a single and two interacting type II excitable units where both the fast and the slow variable are influenced by stochastic perturbations. Our results concern two main issues: $(i)$ determining the $MPAP$s around which the stochastic activation paths are clustered, and $(ii)$ examining in detail the effects of two different noise sources on the statistical features of the activation process, further demonstrating how the statistics is modified due to linear/nonlinear form of interactions. For both issues, we have highlighted the impact of stochastic bifurcation, which underlies the transition from stochastically stable fixed point to stochastically stable limit cycle.

Since the study is focused on the process of first-pulse emission, one of the requirements has been to provide a clear-cut distinction between the spiking responses and the small-amplitude excitations. This has been achieved by introducing an appropriate terminating boundary set, given by the spiking branch of the cubic nullcline. Note that our problem setup is different from the earlier numerical studies on pulse triggering, which have introduced the terminating boundary as a fixed threshold \cite{PPS05,PPM05}, as well as the recent study on a single $FHN$ unit \cite{KPLM13}, where the ghost separatrix has been used as a terminating boundary within the generalized escape problem. The advantages of our approach lie in that the terminating boundary is analytically tractable, while the adopted formulation further facilitates an immediate generalization of the activation problem from the case of a single unit to different scenarios with two interacting units. The differences in the event statistics between the three mentioned problem setups may depend on the system parameters $b$ and $\epsilon$, as well as the noise intensities.

Regarding point $(i)$, it has been established that the topological features of the $MPAP$s qualitatively depend on which type of noise affects the system dynamics. This has been demonstrated by examining the $MPAP$s obtained for three characteristic ratios of external vs internal noise, reflecting the scenarios where a particular noise source or both sources are present. For the fixed characteristic ratio, the $MPAP$ profiles change under increasing noise intensity. The changes become apparent as one approaches the noise values that give rise to stochastic bifurcation. In case of two coupled units, we have shown that the topology of $MPAP$s is substantially affected by the linear/nonlinear form of interactions. For the linear couplings, the respective $MPAP$s of the two units are identical, whereby the solution lies comparably close to that of an uncoupled unit under the analogous parameter set. For the nonlinear couplings, the $MPAP$s of two units may become visibly asymmetrical, depending on the noise intensity. While discussing the topological features of the $MPAP$s, we have also indicated a somewhat surprising numerical finding that the trajectories of the effective set of Hamiltonian equations selected according to a given predefined rule may show striking similarity to the $MPAP$ profiles for scenarios involving both a single and two coupled $FHN$ units. This observation requires a more elaborate study, and suggests that a systematic theory possibly adopting certain elements of the standard Hamiltonian approach to escape problems may potentially be derived for the problem of first pulse emission.

Concerning point $(ii)$, the statistical properties of the first-pulse emission process have been characterized by the dependencies of the mean $TFP$s and the associated coefficients of variation on $D_1$ and $D_2$. Note that the previous work on $FHN$ model has focused on statistics of first exit times in presence of a single noise source \cite{HE05,PPS05,PPM05}, which has typically led to the well-known Kramers result \cite{G04} or its modifications \cite{HE05}. Compared to our approach, the treatment in \cite{PPS05,PPM05} is simplified in terms of definition of the considered quantities and with respect to the terminating boundary conditions. An important novel result is that $\tau$ and $R$ show nearly universal dependence on $D_1$ and $D_2$ for a single unit, as well as for two interacting units. In particular, the mean $TFP$s are found to display three characteristic regimes, whereby the transition from the domain of large $\tau$ values to the plateau region can qualitatively be attributed to the stochastic bifurcation. We have determined the noise intensities that give rise to stochastic bifurcation by introducing the model \eqref{eq13} based on a Gaussian approximation for the dynamics of a stochastic $FHN$ unit. By carrying out the bifurcation analysis of the approximate model, we have obtained the Hopf bifurcation curve $D_2(D_1)$ which qualitatively outlines the stochastic bifurcation of the exact system. In case of two units, the impact of stochastic bifurcation has been found to depend on the form of coupling.

We have further examined the distributions of $TFP$s over an ensemble of different stochastic realizations under fixed $(D_1, D_2)$. So far, little has been known about the profile of corresponding distributions even for a typical escape problem. In fact, the only available analytical result, derived from the theory of large fluctuations, indicates that the exit time distribution for the escape problem asymptotically acquires exponential form \cite{D83}. However, the problem of first pulse emission is conceptually different from the typical escape problem, whereas the noise intensities we consider are small, but finite. Still, in case of a single unit, we find exponential distribution of $TFP$s under prevailing $D_2$, which is consistent with the Poissonian process. However, the activation process dominated by $D_1$ yields a different distribution which shows a unimodal, Lorentzian-like profile. For the interacting units, the profile of $TFP$ distribution is further influenced by the form of coupling.

Apart from the $TFP$ distribution, the form of coupling affects in a nontrivial fashion the correlation of individual activation events. As a qualitative explanation, in subsection \ref{StatCouple} we have suggested a link between the correlation of individual activation events and the synchronization features of the time series for the given parameter set. In this context, the future research should set the ground for potential application of the theory of large fluctuations to the research of stochastic synchronization in excitable systems.

This study has shown how the analysis of activation process may be extended from a single excitable unit to different instances of two interacting units. In the forthcoming paper, our goal will be to examine the noise-driven first pulse emission process in an assembly of excitable units, focusing on whether such an assembly may be considered a macroscopic excitable element, and if so, what may be the analogies and differences compared to the excitable behavior of a single unit.

\begin{acknowledgments}
This work has been supported by the Ministry of Education, Science and Technological Development of the Republic of Serbia, under project No. $171017$. M.P. acknowledges support from the Slovenian Research Agency (Grant P5-0027), and from the Deanship of Scientific Research, King Abdulaziz University (Grant 76-130-35-HiCi).
\end{acknowledgments}


\begin{thebibliography}{99}

\bibitem{I07} E. M. Izhikevich, \emph{Dynamical Systems in Neuroscience: The Geometry of Excitability
and Bursting}, (MIT Press, Cambridge Massachusetts, 2007).

\bibitem{WRK00}{J. White, J. Rubinstein, and A. Kay, Trends in Neurosci. \textbf{23}, 131 (2000).} 

\bibitem{CLC11}{W.-Y. Chiang, P.-Y. Lai, and C. Chan, Phys. Rev. Lett. \textbf{106}, 254102 (2011).} 

\bibitem{YMRBSRL06}{A. Yacomotti, P. Monnier, F. Raineri, B. Ben Bakir, C. Seassal, R. Raj, and J. Levenson,
Phys. Rev. Lett. \textbf{97}, 143904 (2006).} 

\bibitem{LHMY02}{M. A. Larotonda, A. Hnilo, J. Mendez, and A. Yacomotti,
Phys. Rev. A \textbf{65}, 033812 (2002).} 

\bibitem{BYSRBL12}{M. Brunstein, A. M. Yacomotti, I. Sagnes, F. Raineri, L. Bigot, and A. Levenson,
Phys. Rev. A 85, 031803 (2012).} 

\bibitem{LGNS04}{B. Lindner, J. Garcia-Ojalvo, A. Neiman and L. Schimansky-Geier,
Phys. Rep. \textbf{392}, 321 (2004).}

\bibitem{W00} S. Wiggins, \emph{Introduction to Applied Nonlinear Dynamical Systems and Chaos}, 2nd ed.
(Springer, New York, Cambridge, 2000).

\bibitem{DRL12} A. Destexhe and M. Rudolph-Lilith, \emph{Neuronal noise},
(New York, Springer, 2012). 

\bibitem{SJ02}{J. W. Shuai and P. Jung, Phys. Rev. Lett. \textbf{88}, 068102 (2002).}

\bibitem{HE05}{R. C. Hilborn and R. J. Erwin, Phys. Rev. E \textbf{72}, 031112 (2005).}

\bibitem{MS92}{R. S. Maier and D. L. Stein, Phys. Rev. Lett. \textbf{69}, 3691 (1992).} 

\bibitem{MS96}{R. S. Maier and D. L. Stein, J. Stat. Phys. \textbf{83}, 291 (1996).} 

\bibitem{DLMS96}{M. I. Dykman, D. G. Luchinsky, P. V. E. McClintock, and V. N. Smelyanskiy,
Phys. Rev. Lett. \textbf{77}, 5229 (1996).} 

\bibitem{KPLM13}{I. A. Khovanov, A. V. Polovinkin, D. G. Luchinsky, and P. V. E. McClintock,
Phys. Rev. E \textbf{87}, 032116 (2013).}

\bibitem{PPS05}{E. V. Pankratova, A. V. Polovinkin, and B. Spagnolo, Phys. Lett. A \textbf{344}, 43 (2005).}

\bibitem{PPM05}{E. V. Pankratova, A. V. Polovinkin, and E. Mosekilde, Eur. Phys. J. B \textbf{45}, 391 (2005).}

\bibitem{FW12} M. I. Freidlin and A. D. Wentzell, \emph{Random Perturbations of Dynamical Systems},
3rd ed. (Springer-Verlag, Berlin Heidelberg, 2012).

\bibitem{SDG99}{V. N. Smelyanskiy, M. I. Dykman, and B. Golding,
Phys. Rev. Lett. \textbf{82}, 3193 (1999).} 

\bibitem{MS01}{R. S. Maier and D. L. Stein, Phys. Rev. Lett. \textbf{86}, 3942 (2001).} 

\bibitem{BMLSM05}{S. Beri, R. Mannella, D. G. Luchinsky, A. N. Silchenko, and P. V. E. McClintock,
Phys. Rev. E \textbf{72}, 036131 (2005).}

\bibitem{FPTKB14}{I. Franovi\'c, M. Perc, K. Todorovi\'c, S. Kosti\'c and N. Buri\'c,
paper submitted to Phys. Rev. E}

\bibitem{KWS98}{S. Kadar, J. Wang, and K. Showalter, Nature \textbf{391}, 770 (1998).}

\bibitem{KS01}{M. Krupa and P. Szmolyan, SIAM J. Math. Analysis \textbf{33}, 286 (2001).}

\bibitem{BR11}{I. Bashkirtseva and L. Ryashko, Phys. Rev. E \textbf{83}, 061109 (2011).}

\bibitem{DMSSS92}{M. Dykman, P. V. E. McClintock, V. N. Smelyanskiy, N. D. Stein, and N. G. Stocks,
Phys. Rev. Lett. \textbf{68}, 2718 (1992).}

\bibitem{NBK13}{J. M. Newby, P. C. Bressloff, and J. P. Keener, Phys. Rev. Lett. \textbf{111}, 128101 (2013).}

\bibitem{SSSMLM96} {N. G. Stocks, N. D. Stein, H. E. Short, R. Mannella, D. G. Luchinsky, and
P. V. E. McClintock, in \emph{Fluctuations and Order: the New Synthesis}, (Springer, Berlin, 1996),
pp. 53-67}

\bibitem{HSE14}{E. Hunsberger, M. Scott, and C. Eliasmith, Neural Comput. \textbf{26}, 1600 (2014).}

\bibitem{PK97}{A. S. Pikovsky and J. Kurths, Phys. Rev. Lett. \textbf{78}, 775 (1997).} 

\bibitem{DA01} P. Dayan and L. F. Abbott, \emph{Theoretical Neuroscience: Computational and Mathematical
Modeling of Neural Systems}, (MIT Press, Cambridge, 2001). 

\bibitem{A99} L. Arnold, \emph{Random Dynamical Systems}, (Springer Verlag, Berlin, 1999). 

\bibitem{ABR04}{J. A. Acebr\'{o}n, A. R. Bulsara and W.-J. Rappel, Phys. Rev. E \textbf{69}, 026202 (2004).}

\bibitem{GLV09}{M. Gaudreault, F. L\'{e}pine and J. Vi\~{n}als, Phys. Rev. E \textbf{80}, 061920 (2009).} 

\bibitem{GBV11}{M. Gaudreault, J. M. Berbert and J. Vi\~{n}als, Phys. Rev. E \textbf{83}, 011903 (2011).} 

\bibitem{TP01}{S. Tanabe and K. Pakdaman, Phys. Rev. E \textbf{63}, 031911 (2001).}

\bibitem{FTVB13}{I. Franovi\'c, K. Todorovi\'c, N. Vasovi\'c and N. Buri\'c,
Phys. Rev. E \textbf{87}, 012922 (2013).}

\bibitem{BRTV10}{N. Buri\'c, D. Rankovi\'c, K. Todorovi\'c and N. Vasovi\'c,
Physica A \textbf{389}, 3956 (2010).}

\bibitem{LL10} C. Laing and  G. J. Lord, eds., \emph{Stochastic Methods in Neuroscience},
(Oxford University Press, New York, 2010).

\bibitem{G04} C. W. Gardiner, \emph{Handbook of Stochastic Methods for Physics,
Chemistry and the Natural Sciences}, 3rd ed. (Springer-Verlag, Berlin Heidelberg, 2004). 

\bibitem{R89} H. Risken, \emph{The Fokker-Planck Equation}, 2nd ed. (Springer-Verlag, Berlin Heidelberg, 1989).

\bibitem{PB13} W. Paul, J. Baschnagel, \emph{Stochastic Processes: From Physics to Finance}, 2nd ed.
(Springer, Heidelberg, 2013).

\bibitem{D83}{M. V. Day, J. Math. Anal. Appl. \textbf{147}, 134 (1983).}

\bibitem{DVM05}{R. E. Lee DeVille, E. Vanden-Eijnden, and C. B. Muratov,
Phys. Rev. E \textbf{72}, 031105 (2005).}

\bibitem{BJPS09} A. Balanov, N. Janson, D. Postnov, and O. Sosnovtseva, \emph{Synchronization:
From Simple to Complex}, (Springer-Verlag, Berlin Heidelberg, 2009), p. 239-258.

\bibitem{HYPS99}{S. K. Han, T. G. Yim, D. E. Postnov, and O. V. Sosnovtseva, Phys. Rev. Lett.
\textbf{83}, 1771 (1999).}

\bibitem{ZKH03}{C. Zhou, J. Kurths, and B. Hu, Phys. Rev. E \textbf{67}, 030101(R) (2003).}

\bibitem{N94}{A. Neiman, Phys. Rev. E \textbf{49}, 3484 (1994).}

\bibitem{HZ01}{B. Hu and C. Zhou, Phys. Rev. E \textbf{63}, 026201 (2001).}

\bibitem{HL03}{H. Henry and H. Levine, Phys. Rev. E \textbf{68}, 031914 (2003).}

\bibitem{P05}{M. Perc, Phys. Rev. E \textbf{72}, 016207 (2005).}

\bibitem{P07}{M. Perc, Phys. Rev. E \textbf{76}, 066203 (2007).}


\end{thebibliography}
\end{document}